\documentclass[pdflatex,sn-basic]{sn-jnl}
\newif\iftrack \trackfalse 
\usepackage{graphicx}%
\usepackage{multirow}%
\usepackage{amsmath,amssymb,amsfonts}%
\usepackage{amsthm}%
\usepackage{mathrsfs}%
\usepackage[title]{appendix}%
\usepackage{xcolor}%
\usepackage{textcomp}%
\usepackage{manyfoot}%
\usepackage{booktabs}%
\usepackage{algorithm}%
\usepackage{algorithmicx}%
\usepackage{algpseudocode}%
\usepackage{listings}%
\usepackage{color}%
\iftrack
   \usepackage[inline]{trackchanges} 
\else
   \usepackage[finalnew]{trackchanges} 
\fi
\addeditor{LP}
\addeditor{LP Note}
\addeditor{MS}
\addeditor{MS Note}
\addeditor{CP}
\addeditor{CP Note}
\renewcommand{\usercolor}{%
   \ifthenelse{\value{userid} = 0}%
       {\definecolor{UserColor}{rgb}{0.024,0.160,0.380}} 
       {\ifthenelse{\value{userid} = 1}%
         {\definecolor{UserColor}{rgb}{0.160,0.380,0.024}} 
         {\ifthenelse{\value{userid} = 2}%
           {\definecolor{UserColor}{rgb}{0.380,0.024,0.160}} 
           {\ifthenelse{\value{userid} = 3}%
             {\definecolor{UserColor}{rgb}{0.13,0.60,0.65}}%
             {\ifthenelse{\value{userid} = 4}%
               {\definecolor{UserColor}{rgb}{0.86,0.50,0.12}}
               {\ifthenelse{\value{userid} = 5}%
                 {\definecolor{UserColor}{rgb}{0.13,0.70,0.50}}
               }%
             }%
           }%
         }%
       }%
}
\soulregister\ac7     
\soulregister\ref7    
\soulregister\cite7
\soulregister\citeA7
\soulregister\citep7
\soulregister\citepA7
\soulregister\citet7
\soulregister\citetA7
\soulregister\it7
\soulregister\itA7

\newcommand\addlp[1]{\add[LP]{\protect{#1}}}

\newcommand\chalp[2]{\change[LP]{#1 }{\protect \hspace{0.5ex} #2}}
\newcommand\remlp[1]{\remove[LP]{#1}}

\newcommand\addms[1]{\add[MS]{\protect{#1}}}

\newcommand\chams[2]{\change[MS]{#1}{\protect #2}}

\theoremstyle{thmstyleone}%
\theoremstyle{thmstyletwo}%
\theoremstyle{thmstylethree}%
\newcommand{\vhx}{\vspace{0.5ex}}
\newcommand{\vx}{\vspace{1.0ex}}
\newcommand{\Cov}{ \mathop{ \rm Cov }\nolimits }
\newcommand{\tra}[1]{ {#1}^{^{\bf \top}}\! }

\newcommand{\web}[1]{\href{#1}{#1}}
\newcommand{\beq}{ \begin{eqnarray} }
\newcommand{\eeq}[1]{\label{#1}\end{eqnarray}}
\newcommand{\eeqn}{ \nonumber \end{eqnarray} }
\newcommand{\Frac}[2]{\frac{\displaystyle\strut #1}{\displaystyle\strut #2} }
\newcommand{\vex}{\vspace{1ex}}
\newcommand{\dss}{\displaystyle}
\newcommand{\ntab}[2]{ \multicolumn{1}{#1}{#2} }
\newcommand{\nntab}[2]{ \multicolumn{2}{#1}{#2} }

\definecolor{Dred}{rgb}{0.312,0.070,0.070}
\definecolor{Dblue}{rgb}{0.070,0.070,0.312}
\definecolor{Dgreen}{rgb}{0.070,0.312,0.070}
\definecolor{Ugreen}{rgb}{0.070,0.660,0.070}
\definecolor{Db}{rgb}    {0.050,0.0,0.320}

\newcommand{\Number}[1]{\ifnum#1<10\relax0\number#1\else\number#1\fi}
\newcommand{\isodate}{
\count150=\time
\count151=\count150
\divide\count151 by 60
\count151=\count151
\multiply\count151 by 60
\count152=\count150
\advance\count152 by -\count151
\divide\count151 by 60
\count152=\count151
\multiply\count151 by 60
\count153=\count150
\advance\count153 by -\count151
\Number{\year}.\Number{\month}.\Number{\day}--\Number{\count152}:\Number{\count153}
}

\begin{document}

\title[Assessment of EOP accuracy]{Assessment of the Earth orientation parameter accuracy 
      from concurrent VLBI observations}

\author*[1]{\fnm{Leonid} \sur{Petrov}}\email{Leonid.Petrov@nasa.gov}
\author[2]{\fnm{Christian} \sur{Pl\"otz}}\email{Christian.Ploetz@bkg.bund.de}
\author[3]{\fnm{Matthias} \sur{Schartner}}\email{mschartner@ethz.ch}

\received{June 18, 2025}

\affil*[1]{ \orgdiv{Code 61A}, 
            \orgname{NASA Goddard Space Flight Center}, 
            \orgaddress{
                        \street{8800 Greenbelt Rd}, 
                        \city{Greenbelt}, 
                        \postcode{20770}, 
                        \state{MD}, 
                        \country{US}%
                       }%
          }%
\affil[2]{ \orgdiv{Geodetic Observatory Wettzell}, 
           \orgname{Bundesamt f\"ur Kartographie und Geod\"asie}, 
           \orgaddress{
                        \street{Sackenrieder Str. 25}, 
                        \city{Bad K\"otzting}, 
                        \postcode{93444}, 
                        \state{Bayern}, 
                        \country{Germany}%
                       }%
          }%
\affil[3]{ \orgdiv{Institute of Geodesy and Photogrammetry}, 
           \orgname{ETH Zurich}, 
           \orgaddress{
               \street{Robert-Gnehm-Weg 15},
               \city{Zurich}, 
               \postcode{8093}, 
               \state{Zurich}, 
               \country{Switzerland}%
           }%
}
\abstract{
   We have assessed accuracy of estimates of Earth orientation parameters 
(EOP) determined from several very long baseline interferometry (VLBI) 
observing programs that ran concurrently at different networks. We consider 
that the root mean square of differences in EOP estimates derived from 
concurrent observations is a reliable measure of accuracy. We confirmed 
that formal errors based on the assumption that the noise in observables is 
uncorrelated \chalp{are close to useless}{have a limited use}. We found no evidence that advanced 
scheduling strategies with special considerations regarding the ability 
to better solve for atmospheric path in zenith direction applied for 1-hr 
single-baseline sessions have any measurable impact on the accuracy of 
EOP estimates. From this, we conclude that there is a certain limit in
our ability to solve for the atmospheric path delay using microwave
observations themselves and a scheduling strategy is not the factor that 
impairs accuracy of EOP determination. We determined that EOP errors vary 
with season, being smaller in winter and greater in summer. 
\addlp{We got the quantitative estimate of the impact of unmodeled source 
structure on EOP estimates and we found that the seasonal extra variance 
is one order of magnitude greater than the impact of source structure.} 
We \chalp{found}{established} that the EOP errors are scaled with an increase in duration of 
an observing session as a broken power law with the power of -0.3 at 
durations longer than 2--4 hours, which we explain as a manifestation of 
the presence of correlations in the atmospheric noise.
}


\maketitle


\section{Introduction}\label{s:intro}

  It was demonstrated in the 1980s that Very Long Baseline Interferometry
(VLBI) outperforms traditional optical techniques for determination of 
irregularities in the Earth rotation by more than an order of magnitude 
\citep{r:merit82,r:merit00}. Regular VLBI observations dedicated to 
the determination of the Earth orientation parameters (EOPs) commenced on 
January 04, 1984 and are running since then without interruption. In order 
to provide a meaningful interpretation to a given parameter adjusted from 
measurements, its error should be characterized. It turned out to be 
a difficult task to assess the accuracy of the EOP from VLBI observations.
VLBI does not directly measure Earth rotation. It measures voltage at 
receivers, which is the thermal noise with a tiny admixture of the noise from 
an observed extragalactic radio source. Lengthy records of voltage data --- 
the dataset that we process for this study has over $3 \cdot 10^{17}$ 
samples, undergo several stages of data analysis, including non-linear 
procedures. The last stage uses group delays as an input and estimates EOP 
using least squares. If we know precisely the covariance matrix of noise 
in group delays, we can easily arrive at dispersions of EOP estimates using 
the classical error propagation law. We do not know. As a coarse 
approximation, we form diagonal terms of the input covariance matrix from 
estimates of uncertainties of group delays {\it setting off-diagonal terms 
to zero}. It was known since 1980s \citep{r:her83,r:ray91} that 
uncertainties of estimates of EOP and any other parameters derived that way
are underestimated in a wide range from a factor of 1.2 to 10. These 
estimates of uncertainties are traditionally labeled as ``formal'' 
staying short from telling that these estimates of uncertainties are just 
wrong.

  Because of a deficiency of formal errors derived with the use of a~priori
covariance matrix with off-diagonal terms set to zero, we have to seek
alternative ways to assess the accuracy of our results. Since no other 
technique can estimate Earth's spin angle more precisely than VLBI, we 
have to compare VLBI results against other results from independent VLBI 
experiments. Since the Earth orientation strongly varies with time, we 
cannot compare parameters determined from one epoch to parameters 
determined from another epoch. Thus, we arrived at a concept of concurrent 
independent observations, whose analysis has the potential 
to provide a metric of EOP accuracy more robust than formal uncertainties. 
In this work, we will provide results of our re-analysis of group delays 
of three dedicated VLBI observing campaigns in 1997.0--2000.5, in 
2019--2024.5, in 2022.3--2024.7, and several other campaigns that 
overlapped with them in time. The goal of our study is to evaluate 
the accuracy of EOP determined from these concurrent observations 
and investigate which factors affect it.

\section{Formalism of the Earth orientation parameters}\label{s:for}

  Since space is three-dimensional, a rotation of a rigid body is 
described by three independent parameters. \chalp{The classical 
formalism of rotational kinematic was established by Leonhard Euler, 
and these rotation angles bear his name.}{We follow the classical
approach for describing rotation of a rigid body established in the
XVIII century and that is introduced in modern physics textbooks
\citep[see, for instance,][]{r:Goldstein_book,r:Taylor_book,r:Landau_Lifshitz_book}.
In the Cartesian coordinate system a transformation from 
the inertial space to the body-fixed coordinate system is 
described by three Euler angles.}
Considering that we estimate a small perturbation of 
Euler angles with respect to an a~priori model, we can linearize 
equations of motion and write the coordinate transformation of
a vector $\vec{r}_{{}_T}$ from the crust-fixed terrestrial 
coordinate system to the vector in the inertial celestial coordinate 
system $\vec{r}_{{}_C}$ in a form
\beq
   \vec{r}_{{}_C} = \widehat{\mathstrut\cal M}_a(t) \, \vec{r}_{{}_T} 
                    \: + \:
                    \vec{q}_e(t) \times 
                    \vec{r}_{{}_T},
\eeq{e:e1}
   where  $\widehat{\mathstrut\cal M}_a(t)$ is the a~priori matrix 
of the Earth rotation and $\vec{q}_e(t)$ is the \chalp{vector}
{right-handed 3-vector in the Cartesian orthogonal basis} of small residual
Euler angles with components that we will label as $E_1$, $E_2$, $E_3$. 
\addlp{Since this approach is based in classical physics textbooks, we call
it classical}. Historically, polar motion and Universal Time were determined 
with two totally different techniques. Polar motion was determined from 
observations of zenith telescopes with a micrometer that was graduated 
with arcseconds. Therefore, \chalp{a natural}{an instrumental} unit of polar 
motion was arcsecond. Universal Time was determined by measuring the time 
of passing a star through the meridian with respect to a local clock using 
the eye-hear method or, later, using photoelectric registration. Therefore, 
\chalp{a natural}{an instrumental} unit of Universal Time was second. Neither 
method has been used for decades. Internally, partial derivatives of VLBI 
time delay with respect to components of the Euler rotation vector are 
written \addlp{and coded} in radians. Therefore, natural units of EOP 
determined with VLBI are radians. \addlp{We use radians in our work because
this the SI standard unit for angle and because it is essential for our
work to use the same units for all three components of the Earth 
rotation for their intercomparison.}
\remlp{
Since 
we will compare $E_1$, $E_2$, and $E_3$, we found it is extremely 
inconvenient to use different units for components of the same vector 
and apply artificial scaling factors. Since we have to change unit 
scaling factors anyway, we set all of them to one and adhere SI unit 
for angles in our work, namely, radian. For a reference: 
1~nrad = 0.206~mas = 13.8~$\mu$sec. Euler angles are related to vintage 
polar motion $X_p$, $Y_p$, and UT1 parameters as
}
%
\remlp{
   where $X_p$ and $Y_p$ are expressed in arcseconds, UT1 in seconds,
and scaling factors 
$\mu = \pi/(180 \cdot 60 \cdot 60) \approx 4.848 \cdot 10^{-6}$, 
$ \kappa = -\pi/(12 \cdot 60 \cdot 60)  \cdot (\Omega_n + \Omega_p)/\Omega_n 
\approx -7.292115 \cdot 10^{-5}$. Here $\Omega_n$ is the nominal Earth 
rotation angular velocity and $\Omega_p$ is the precession rate in 
right ascension.
}

\remlp{
  Although the fact that space is three-dimensional seems firmly 
established, it is not uncommon to encounter a statement that there are 
five parameters that describe Earth rotation, for instance, in 
It was \citet{r:her86a} who suggested estimating daily nutation offsets 
in the mid-1980s as a temporary measure because by that time the time 
series of VLBI experiments was too short to separate the principal 
nutation term with a period of 18.6~years, annual term, and precession. 
Nutation can be considered as a component of quasi-diurnal harmonic 
oscillations in the polar motion that occupies a certain part of the 
spectrum. Nowadays, the time span of high-quality VLBI observations is 
sufficient for the separation of nutation constituents, and we are 
in a position to return to a classical approach for estimation of 
the nutation term amplitudes in the frequency domain set by James 
Bradley in 1748 \citep[see][]{r:bradley} as it has been demonstrated 
by \citet{r:erm} and now routinely used for common VLBI analysis 
\citep{r:rfc1}. The high-frequency spectrum of the Earth rotation 
exhibits a minor variability due to the free core nutation, and this 
variability causes estimates of the EOP spectrum to evolve with 
an increase in the time interval used for estimation. However, since 
the free-core nutation has a period of around 433 days, this variability 
is slow, and it is sufficient to re-evaluate the spectrum of 
high-frequency EOP every 2--5 months to capture this signal using 
a global dataset, including newly acquired data. The anharmonic 
constituent of EOP is much more variable and has to be estimated over 
a much shorter period of time, 0.5--5 days, using datasets that cover 
these intervals.
}

\remlp{
  In the context of our work, we consider harmonic variations in the 
Earth's rotation as constant for a given dataset, but Euler angles as 
freely varying with time. We restrict the scope of our analysis to 
three Euler angles estimated over intervals of 0.5 to 24 hours long.
}
\addlp{The Earth orientation parameters in the notation that is 
currently used by the International Earth Rotation Service (IERS) 
can be converted to the classical Euler angles notation as

\beq
   \begin{array}{lcl}
      E_1(t) & = & \mu \, Y_p(t) \vspace{0.5ex} \\
      E_2(t) & = & \mu \, X_p(t) \vspace{0.5ex} \\
      E_3(t) & = & \kappa  \, \rm (UT1-UTC)(t), \\
   \end{array}
\eeq{e:e2}
}
   where $X_p$ and $Y_p$ are expressed in arcseconds, UT1 in seconds,
and scaling factors 
$\mu = \pi/(180 \cdot 60 \cdot 60) \approx 4.848 \cdot 10^{-6}$, 
$ \kappa = -\pi/(12 \cdot 60 \cdot 60)  \cdot (\Omega_n + \Omega_p)/\Omega_n 
\approx -7.292115 \cdot 10^{-5}$~rad/s. Here $\Omega_n$ is the nominal Earth 
rotation angular velocity and $\Omega_p \approx 7.086183 \cdot 10^{-12}$~rad/s 
is the precession rate in right ascension. 
For a reference: 1~nrad = 206~${\mu}$as = 13.8~$\mu$sec. 

  The vector of perturbations of Euler angles $\vec{q}_e$ is 
a smooth function of time. We can approximate it using simple analytical
functions that depend on some numbers collectively known as Earth
Orientation Parameters. A search of optimal parameterization is 
an important problem of geodesy \citep[see, for instance,][]{r:biz23}. 
Evaluation of harmonic variations from observations poses a certain difficulty. 
Historically, for 250 years amplitudes of the spectrum of harmonic variations 
of $E_1, E_2$ in the retrograde diurnal band called nutations were estimated 
from observations. It was \citet{r:her86a} who suggested estimating daily 
nutation offsets in the mid-1980s as a temporary measure to overcome 
degeneracy because at that time the time series of VLBI experiments was too 
short to separate the principal nutation term with a period of 18.6~years, 
the annual term, and precession rate. This approach is widely used and became 
rather common \citep{r:iers2010}. Nowadays, the time span of high-quality 
VLBI observations is sufficient for the separation of nutation constituents, 
and we are in a position to return to a classical approach for estimation 
of the nutation term amplitudes in the frequency domain set by James 
Bradley in 1748 \citep[see][]{r:bradley} as it has been demonstrated 
by \citet{r:erm} and nowadays routinely used for common VLBI analysis 
\citep[see, for instance,][]{r:jmg96,r:ibnu20,r:rfc1}. The detailed 
discussion of advantages and disadvantages of the parameterization of 
harmonic variations in the Earth rotation in a form of nutation daily 
offsets or in a form of complex amplitudes of the spectra. The 
transformation between two parameterizations can be found in \citet{r:erm} 
\addlp{and a brief summary is given in appendix~\ref{a:hist_eop}.}

  In the context of our work, we consider harmonic variations in the 
Earth's rotation that we solve for in our analysis as constant for 
a given dataset, but residual anharmonic Euler angles as freely varying 
with time. We restrict the scope of our analysis to three Euler angles 
estimated over intervals of 0.5 to 24 hours long.

\section{Observations}\label{s:obs}

  We used all 24-hr geodetic multi-baseline VLBI data since 1980 
April 11 through 2025 March 03 in our analysis, in total 7818 
experiments. 

  In addition, we processed data from a number of 1-hr 
single baseline geodetic campaigns\footnote{A small number of 
experiments of these campaigns had more than two stations.}. 
We split these data into several subsets of target data, reference
data, and remaining data labeled as background geodetic data. 
We process them as well, but we report in this study only 
results of EOP determination from target experiments.

  We designated the following multi-baseline 24-hour experiments 
as the target data:

\begin{itemize}

   \item 148 24-hr experiments of regular quad-band programs VGOS-OPS, 
         VGOS-TEST, and VGOS-RD, collectively called in this study vo, 
         that ran in 2019.0--2024.0. In these experiments antennas 
         of 12--13~m diameter with the slewing speed up to $12^\circ/s$ 
         dubbed as VLBI Global Observing System \citep[VGOS;][]{r:nie18}
         were used. Fast slewing speed greatly simplifies schedule 
         optimization, in particular, allowing for an increased 
         distribution of observations among azimuth and elevation angles. 
         The use of four bands spread over the 7~GHz bandwidth allows 
         determination of group delay with a reported precision of 
         3--5~ps. The network included 13~stations. It varied from 
         3 to 11~stations in a given experiment, with a median of 
         7~stations.

   \item 10 quad-band experiments at stations Mg and Ws under program 
         VGOS-24INT-S, for brevity called s22, in 2022.3--2024.7. These 
         experiments were scheduled as 22~consecutive 1-hr experiments 
         using two different strategies and then concatenated. Since Ws 
         participated in daily operational experiments under another 
         program, the total session duration was 22~hours.

   \item 81 dual-band 24-hr experiments from program CORE-A, thereafter 
         contracted to ca in 1997.1--2000.5 
         \citep{r:core99,r:core01,r:core02} at a global  network of 
         13 stations. It varied from 4 to 6 stations in a given experiment
         with the median of 6 stations.
         
   \item 235 dual-band 24-hr experiments from concurrent NEOS-RAPID, 
         IVS-R1, IVS-R4 programs \citep{r:r1r4} have been running since 
         1993.4 till present and are concurrent with one of the 
         above-mentioned programs. For brevity, we contract program names 
         to na and r1r4. Two 24-hr experiments are considered concurrent 
         if the middle epochs of these experiments are within 1 hour. 
         The network included 32~stations. It varied from 4 to 15~stations 
         in a given experiment, with the median of 8~stations.
\end{itemize}

\iftrack \newpage \fi 

   We also considered the following single-baseline 1-hr experiments as 
additional targets:

\begin{itemize}

   \item 3275 dual-band 1-hr experiments at stations Kk and Wz from 
         IVS-INT-1 and IVS-INT-00 program in 1999.9--2025.2. Observations at 
         this baseline run \chalp{3--4}{3--5} days a week. We call this program KkWz.

   \item 423 quad-band 1-hr experiments at stations K2 and Ws from VGOS-INT-A
         program in 2020.2--2025.2. Observations at this baseline run \chalp{3--4}{3--5}
         days a week. We call this program K2Ws.

   \item 133 dual-band 1-hr experiments at stations Ts and Wz from IVS-INT-2
         program in 2002.8--2017.0. Observations at this baseline run 
         2--3 days a week till the \addlp{Ts} antenna was decommissioned. 
         We call this program TsWz. \addlp{Since 2023 observations between 
         Wz and station Is that is located close to Ts resumed. We call that 
         program IsWz, and we included 65 such experiments.}

   \item 115 dual-band 1-hr experiments at stations Mk and Wz from IVS-INT-2
         program in 2019.1--2025.2. Observations at this baseline run 
         2--3 days a week. We call this program MkWz.

   \item 58 dual-band 1-hr experiments at stations Hn and Wz from USNO-INT-P
         program.
\end{itemize}

\begin{figure}
    \centering
    \includegraphics[width=0.99\linewidth]{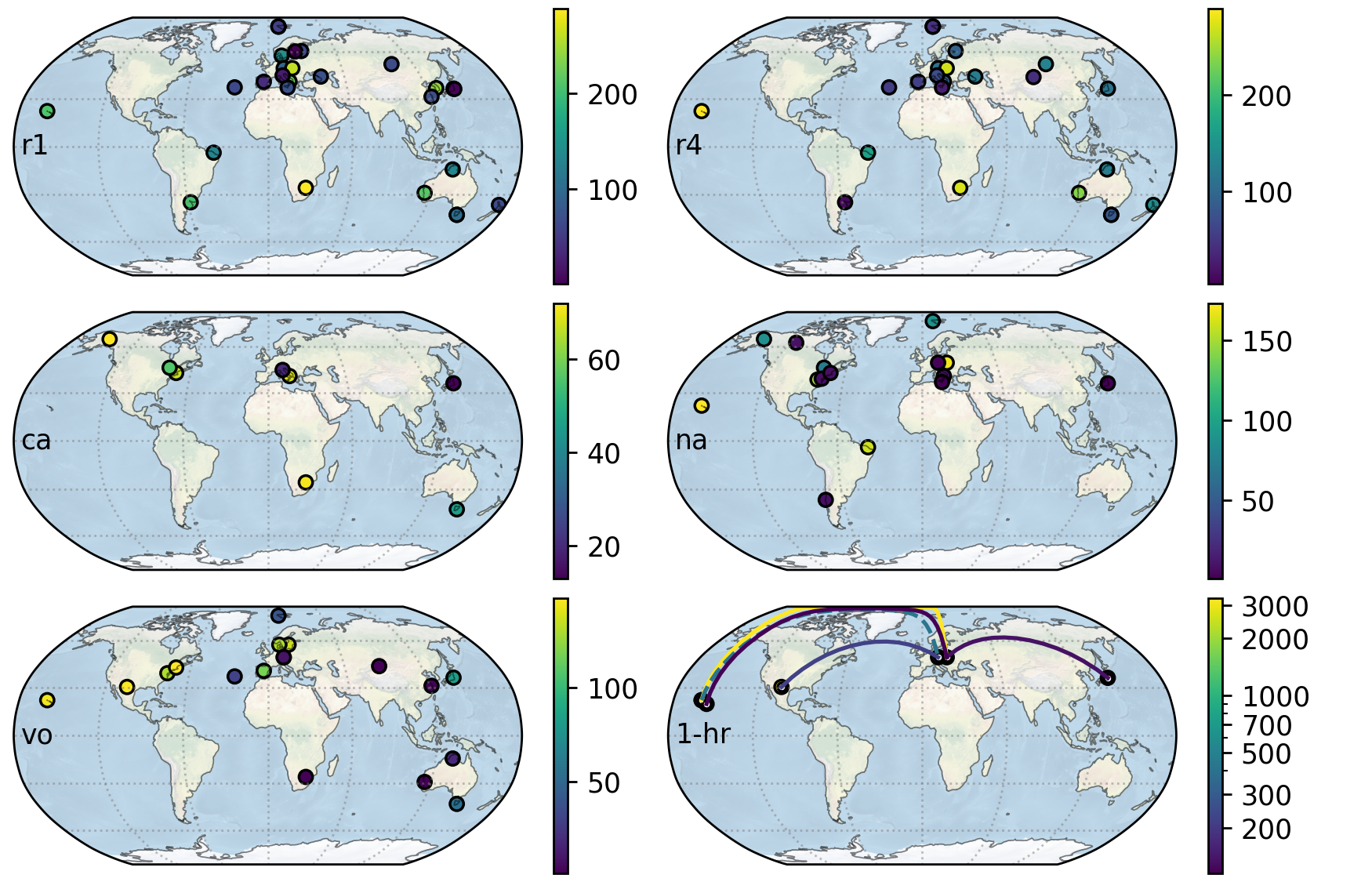}

    \caption{Station networks for different observing programs, 
             color-coded by the number of experiments used in our
             analysis. Baseline from 1-hr experiments are explicitly 
             depicted. Close telescopes (twins) are slightly offset
             in the maps.
            }
    \label{f:network}
\end{figure}

   All these experiments ran concurrently with 24-hr experiments from
r1r4 program. Positions of stations from 1-hr programs are presented in 
Table~\ref{t:1hr_sta} and their location is shown in Figure~\ref{f:network}.

\begin{table}
   \caption{Positions of stations participated in 1-hr single-baseline
            experiments.}
   \begin{tabular}{llrrr}
       \hline \\
       Name & Full name & \ntab{c}{$\phi_{\rm geod}$} 
                        & \ntab{c}{$\lambda$} 
                        & \ntab{c}{$h_{\rm ort}$} \\
            &           & \ntab{c}{deg} 
                        & \ntab{c}{deg} 
                        & \ntab{c}{m} \\
       \hline \\
       Hn  & HN-VLBA   & 42.9336 & 288.0134  &  295.6 \\
       K2  & KOKEE12M  & 22.1264 & 200.3351  & 1151.6 \\
       Kk  & KOKEE     & 22.1266 & 200.3349  & 1160.0 \\
       Mg  & MACGO12M  & 30.6803 & 255.9763  & 1912.7 \\
       Mk  & MK-VLBA   & 19.8014 & 204.5445  & 3763.1 \\
       Ts  & TSUKUB32  & 36.1031 & 140.0887  &   44.7 \\
       Ws  & WETTZ13S  & 49.1434 &  12.8783  &  625.7 \\
       Wz  & WETTZELL  & 49.1450 &  12.8775  &  622.3 \\
       \hline \\
   \end{tabular}
   \label{t:1hr_sta}
\end{table}

   The experiments listed above are not the only concurrent geodetic 
observations. There were other concurrent programs, but they either 
covered a short interval of time, for instance, a 15-day long CONT-17 
program \citep{r:cont17_org}, or had fewer experiments. Since 
differences in statistics of various programs are expected to differ 
at a level of tens of percents, a significant dataset needs to be 
analyzed in order to make an inference about whether the difference 
in statistics are significant or not. For instance, to establish that 
the standard deviations of two samples that differ at 20\% are 
significant at a~0.05~level, the F-test \citep{r:fisher25} requires 
a sample size of at least 118.

\subsection{Scheduling S22 campaign}\label{s:sched_22}

  During scheduling VLBI Intensive sessions, several criteria are 
usually taken into account to be optimized \citep{r:vips++a}. 
These include:
(1) maximizing the total number of observations,
(2) obtaining observations at a range of elevation angles 
to enable accurate estimation of atmospheric path delays
in zenith direction,
(3) ensuring a balanced distribution of observations across 
multiple sources, and
(4) specifically including sources at low declinations and 
near the cusps of the mutual sky visibility \citep{r:scha21}.

\addlp{We used two strategies for scheduling in each experiment: 
strategy A for segments with odd indices and strategy B for segments
with even indices. They are described in detail in 
appendix~\ref{a:sched_22}.}

\iftrack \newpage \fi 

\remlp{
  While generally speaking, observations at longer baselines are more 
sensitive for measuring Earth's rotation angle $E_3$, they also face 
limitations due to a reduced common visibility of the sky
\citep{r:scha21}. The $\sim\! 8400$~km baseline between stations Mg and 
Ws presents certain geometric advantages. 
At this distance, observations near the zenith at one station correspond 
to observations at very low elevation angles at the other station 
(see Figure~\ref{f:scheduling_el}). This geometry is particularly 
beneficial for solving for residual atmospheric path delay in zenith
direction, which remains the primary source of error in VLBI measurements. 
To effectively separate atmospheric zenith path delay from other 
parameters, such as station height and clock biases, it is essential 
to observe at a wide range of elevation angles.
%

  The s22 sessions were specifically designed to investigate improved
estimation of residual atmospheric path delay in zenith direction.
Each session is divided into 1-hour blocks that alternate between two 
scheduling strategies. Strategy~A follows the conventional scheduling 
approach utilized in the VieSched++ scheduling software 
\citep{r:scha19} and serves as a baseline reference. By contrast, 
strategy~B is tailored to maximize the variability in elevation angles, 
promoting frequent alternation between high- and low-elevation 
observations to improve sensitivity to zenith atmospheric path delay 
parameters.

  To implement Strategy~B, sources at high elevation for each station are 
identified in advance. Scans are then scheduled according to the following 
repeating pattern: 
(1) a source at high elevation for Mg (corresponds to low elevation for Ws),
(2) any source,
(3) a source at high elevation for Ws (corresponds to low elevation for Mg),
(4) any source.

  The interleaved ``any source'' steps (2 and 4) help to maintain a balanced 
distribution of observations across different sources. In general, a minimum 
interval of 20 minutes is enforced between consecutive observations of the 
same source. However, due to the limited availability of high-elevation sources 
at each station, this restriction is relaxed to 10 minutes for those specific 
sources in strategy B. Additionally, to ensure adequate coverage of 
low-declination regions, a low-declination source is explicitly scheduled 
every 15 minutes in both strategies.
}

\section{Data analysis}\label{s:da}


  \addlp{Conventional geodetic instruments like theodolites or laser
range meters perform direct observations, and assessment of errors of such
individual observation is intuitively clear. VLBI techniques does not measure
Earth rotation directly, and the error propagation through a very complicated
pipeline is rather obscure.}

  The first active signal amplification of the voltage at a receiver at 
radio frequency bands is made by the cryogenically cooled radio telescope's 
front-end system. These radio frequency voltage signals of interest are 
typically down-converted to lower radio frequency (RF) bands for further 
handling. The data acquisition terminal typically comprises a \addlp{2-bit} 
digitizer for RF bands as a backend located in the operations room, which 
converts the analog voltage signal into a time and amplitude discrete digital 
representation of the frequency band-limited analog signal. These digital 
converter units \addlp{governed by the Hydrogen maser} extract a determined 
number of individual intermediate frequencies (IF) from each digitized 
RF-frequency band and convert them into the baseband \chalp{channels.}
{steams that are written alongside with time stamps.}
\remlp{The individual channels are grouped 
as a set of channels together, parameterized by start frequency, bandwidth, and 
amplitude resolution. They are digitally formatted, time-tagged, and then 
recorded on a data storage for further processing. For example, a 
quad-band setup has 32~channels with 32~MHz channel bandwidth and 2-bit
amplitude resolution for one linear polarization of the received radio 
wave. Thus, for both linear polarizations, horizontal and vertical, 
64 streams of formatted data are recorded at a rate of 8 Gigisamples/s. 
All signal operations, from radio frequency conversion downstream 
to individual digital baseband channels, are coherently driven by the 
reference frequency of the Hydrogen maser and inherit its frequency 
stability.}

 \addlp{Contribution of emission from an observed radio source
and thermal emission of the receiver itself are very close to the white
noise, and the root mean square (rms) of the latter noise is typically 
three to five orders of magnitudes greater than the rms of the emission 
from the radio source.}

  Streams of raw voltages formatted samples are processed with the DiFX 
software correlator \citep{r:difx1,r:difx2}. It reads raw voltage
samples, unpacks them from 2-bit amplitude sampling to a floating-point 
representation, shifts the data streams according to a coarse path delay
model to some arbitrary reference points, performs Fourier transform, 
computes complex auto- and cross-spectra, averages them with time, and 
then writes the results that are called visibilities. \remlp{In addition,
time series with amplitude and phase of the phase calibration tones are 
also written into individual files for each scan and station.} Results 
of this correlation processing are called Level 1A 
data\footnote{See \web{https://github.com/nasa/sgdass/blob/main/pima/doc/swin\_format.pdf}
for specifications.}.

   Using visibility data, we evaluated residual phase and group delay
$\tau_p$ and $\tau_g$, as well as its time derivative $\dot{\tau}_p$,
using fringe fitting software fourfit\footnote{See 
\web{https://www.haystack.mit.edu/wp-content/uploads/2021/01/vgos-data-processing.pdf}
accessed on May 16, 2025.}. It combines various algorithms 
to calibrate for the complex bandpass of the data acquisition system and
to derive group delays and delay rates using a non-linear optimization 
for each observation. The estimated residual group delays and phase 
delay rates are sought that when they are applied to phases of the 
complex cross-spectra time series $c_{kj}(f,t)$, the coherent sum over 
all spectral channels and all time epochs of a given observation
reaches the maximum: 

\beq
   \hspace{-1em}
     C(\tau_p,\tau_g,\dot{\tau}_p) e^{-i2\pi \, f_0 \tau_p} = &
          \dss \sum_k \sum_j c_{kj} \times
                e^{i 2\pi( f_0  \dot{\tau}_p (t_k-t_0) \; + \;
                      (f_j - f_0) \tau_g )}.
\eeq{e:e3}

\addlp{Since thermal noise from two datastreams is independent, but
the common emission from an observed source is the same, averaging reduces
the impact of the thermal noise to $C(\tau_p,\tau_g,\dot{\tau}_p)$.}

   In a case of quad-band data, the fringe fitting process estimates phase, 
ionosphere-free group delay, and the contribution of the ionosphere in 
a single fit. In a case of dual-band data, we process visibilities
from each band separately, and then form a linear combination of group 
delays with coefficients selected in such a way that the frequency-dependent 
contribution of path delay in the ionosphere is canceled 
\citep[see, for instance,][]{r:sba}.

   Fourfit computes the uncertainties of group delays based on the 
achieved signal to noise ratio (SNR) of \addlp{$C(\tau_p,\tau_g,\dot{\tau}_p)$  
amplitudes} assuming individual samples of visibilities are independent and 
uncorrelated with time and frequency using the equations 6.45 and A12.24
presented in \citet{r:tms}:

\beq
     \sigma_{\rm gr} = \Frac{1}{2\pi \, \rm SNR} \;
                       \Frac{1}{\Delta f_{\rm rms}},
\eeq{e:e4}
  where $\Delta f_{\rm rms}$ is the root mean square of IF within the band.
\addlp{We have to stress that $\sigma_{\rm gr}$ in expression~\ref{e:e4}
is based on the assumption that the noise process affects phase and amplitude 
of $C$ equally. This assumption is valid for specific noise processes. For 
instance, the delay uncertainly inversely depends on the rms of receiver 
thermal noise. However, processes like delay in the propagation media and 
signal chain affect amplitude in a much lesser extent than phase. If phase 
variations during an observation are $\Delta \phi \gg  2\pi$, then their 
impact on group delay  delay is proportional to $\Delta \phi^{-1}$, but the 
impact on amplitude is proportional to $\Delta \phi^{-2}$. Therefore, group delay 
uncertainty computed using expression~\ref{e:e4} does not fully account the 
contribution of the phase noise, unless it causes phase variations during 
an observations comparable to a phase turn. This is one of the factors that
eventually leads to an underestimation of EOP estimate errors.}

  The ionosphere-free group delay or a combination of group delays 
\chalp{is}{and their uncertainties are} the basis of further data analysis.

\subsection{General geodetic data analysis}

  We estimated parameters in a single least squares run using all
geodetic data accumulated since 1980 April 11 through 2025 March 03,
in total 19.8 million ionosphere-free group delay observables.
Geodetic analysis is reduced to the computation of the theoretical model,
forming differences between observed and theoretical delay combinations,
defining the list of estimated parameters, computation of partial 
derivatives of group delay over these parameters, running an iterative
preliminary least squares solution with outlier elimination,
update of input weights based on reported group delay uncertainties
by evaluation of experiment- and baseline-dependent variances that,
being added in quadrature, make the ratios of weighted post-fit 
residuals to their mathematical expectation close to unity, 
and running a final single least squares solution.

  The general VLBI data analysis procedure is presented in full detail in 
\citet{r:rfc1}. To avoid unnecessary repetition, we present here only
highlights. We applied 3D displacements due to ocean tidal, ocean 
non-tidal, atmospheric pressure, and land water storage loadings 
provided by the International Mass Loading Service \citep{r:malo15}
and computed from the outputs of NASA Global Modeling Office numerical
weather models. We applied in data reduction a~priori slant path 
delays computed by a direct integration of equations of wave propagation 
through the heterogeneous atmosphere \citep{r:padel} using the output of 
NASA numerical weather model GEOS-IT which is a successor of the older 
model GEOS-FPIT \citep{r:reichle11,r:geos12,r:geos18}.

  The total number of estimated  parameters in the final solution exceeds 
5~millions. To invert a matrix of that size, we used a three-level
partitioning: global parameters that were estimated using the entire 
dataset, local parameters that were estimated for each observing session, 
and segmented parameters that were estimated for each station for 
time interval that is shorter than an observing session. 
 
  We estimated the following global parameters: 1)~positions of all the 
stations at the reference epoch 2000.01.01; 2)~linear velocities of all 
the stations; 3)~antenna axis offsets of most of the stations; 4)~sine 
and cosine components of harmonic site position variations of 69 stations 
with a long history of observations at diurnal, semi-diurnal, annual, and 
semi-annual frequencies; 5)~B-spline coefficients that model the 
non-linear motion of 29 stations that included sudden co-seismic position 
changes, smooth post-seismic relaxation, non-linear change of the 
antenna tilt, non-linear uplift due to glaciers melting, non-linear local 
motions, and discontinuities due to station repair; 6)~coefficients of 
harmonic variations of the EOP vector $P_c, P_s, S_c, S_s, R_c, R_s$ at 
890 frequencies including nutations; and 7)~positions of all the sources 
with at least three usable observations.

  When processing  24-hr experiments, we estimated Euler angles 
\remlp{at the specified epoch and their time derivatives as local parameters.}
\addlp{
$E_{1s}, E_{2s}, E_{3s}$ at the specified epoch $t_{\rm ref}$ and their 
time derivatives as local parameters:

{\small
\beq
    \vec{q}_e(t) = \left(
    \begin{array}{ l@{\;} l@{\;} l l@{\;} l }
       E_{1}(s) + \dot{E}_1(s) \cdot (t - t_{\rm ref})  &  + &
       \displaystyle\sum_{j=1}^{N} \left( P^c_{j} \cos (\omega_j \, t + \phi_j) \: + \:
                                          P^s_{j} \sin (\omega_j \, t + \phi_j) \right) +  \\ 
       & & (t - t_0) \, \left( S^c \cos (- \Omega_n \, t + \phi_p)\: + \:
                         S^s \sin (- \Omega_n \, t + \phi_p)  \right) 
       \vspace{2.5ex} \\
       E_{2}(s) + \dot{E}_2(s) \cdot (t - t_{\rm ref}) &  + & 
        \displaystyle\sum_{j=1}^{N} \left( P^c_{j} \sin (\omega_j \, t + \phi_j)\: - \:
                                           P^s_{j} \cos (\omega_j \, t + \phi_j) \right) + \\
      & & (t - t_0) \, \left( S^c \sin ( - \Omega_n \, t + \phi_p) \: - \:
                         S^s \cos (- \Omega_n \, t + \phi_p) \right)
      \vspace{2.5ex} \\
       E_{3}(s) + \dot{E}_3(s) \cdot (t - t_{\rm ref}) &  + &
       \displaystyle\sum_{j=1}^{N} \left( R^c_{j} \cos (\omega_j \, t + \phi_j) +  
                                          R^s_{j} \sin (\omega_j \, t + \phi)j) \right) 
       \\
    \end{array}
    \right),
\eeq{e:e43}
}
  as well as global parameters mentioned above. Here we added symbol $(s)$ 
to emphasize that these parameters do not depend on time, but on the 
specific observing session $s$. Relationship between estimated parameters
$E_1(s), E_2(s), E_3(s)$ and parameters $X_p, Y_p,$ \rm UT1-UTC
used by IERS are given in appendix~\ref{a:hist_eop}.
}

  We also estimated clock breaks and baseline-dependent clock for some experiments 
with clock function discontinuities or baseline-dependent biases. 
In addition, we estimated B-spline coefficients of the 1st degree 
for residual atmospheric path delay in the zenith direction, tilts of the 
symmetry axis of the refractivity field, also known as atmospheric 
gradients, and clock functions for all the stations, except the one 
taken as a reference, as segmented parameters. Time spans of 
B-splines were 20~minutes for atmospheric zenith path delay, 6~hours 
for tilts, and 60~minutes for clock function. 

  When processing  1-hr experiments, \addlp{in addition to global parameters},
we estimated Euler angle $E_3$ as a local parameter, clock function and residual 
atmospheric path delay as segmented parameters:
\addlp{
{\small
\beq
     \vec{q}_e(t) = \left(
    \begin{array}{ l@{\;} l@{\;} l l@{\;} l }
       \phantom{E_{3}(s)} &  &
       \displaystyle\sum_{j=1}^{N} \left( P^c_{j} \cos (\omega_j \, t + \phi_j) \: + \:
                                          P^s_{j} \sin (\omega_j \, t + \phi_j) \right) +  \\ 
       & & (t - t_0) \, \left( S^c \cos (- \Omega_n \, t + \phi_p)\: + \:
                         S^s \sin (- \Omega_n \, t + \phi_p)  \right) 
       \vspace{2.5ex} \\
       \phantom{E_{3}(s)} &  & 
        \displaystyle\sum_{j=1}^{N} \left( P^c_{j} \sin (\omega_j \, t + \phi_j)\: - \:
                                           P^s_{j} \cos (\omega_j \, t + \phi_j) \right) + \\
      & & (t - t_0) \, \left( S^c \sin (- \Omega_n \, t + \phi_p) \: - \:
                              S^s \cos (- \Omega_n \, t + \phi_p) \right)
      \vspace{2.5ex} \\
       E_{3}(s) &  + &
       \displaystyle\sum_{j=1}^{N} \left( R^c_{j} \cos (\omega_j \, t + \phi_j) +  
                                          R^s_{j} \sin (\omega_j \, t + \phi_j) \right) 
       \\
     \end{array}
     \right),
\eeq{e:e46}
}
  as well as global parameters, and kept $E_1, E_2$ fixed.
}
Time spans of B-splines were 10~minutes for atmospheric zenith path delay 
and 20~minutes for clock function. \addlp{We used this
relatively short time spans because we identify the presence of 
the atmospheric signal in residuals of some experiments when we used longer 
time spans.}

  We also imposed constraints on estimated parameters. An exhaustive 
description of constraints can be found in \citet{r:rfc1}. Here we 
stress that Earth Orientation Parameters were unconstrained, and only
minimum constraints were imposed on the station positions and their 
derivatives. The motivation for our choice of running global solutions 
with estimating all parameters, including EOP, in a single least 
squares run, rather than fixing some parameters, is that this
automatically imposed consistency of the estimated station 
positions and EOP. \addlp{Estimation of nutation daily offset from
processing 1-hr experiments is not feasible.} It was reported in the 
past that fixing nutation, polar motion, or station positions to some 
a~priori values that are considered ``good'' resulted in a measurable 
degradation of statistics because of inconsistency that a~prior 
models \citep{r:not08,r:dieck23}. \addlp{Our approach of estimating
harmonic EOP variations as global parameters overcomes this limitation.}
In addition, by solving for station positions and source coordinates,
we effectively incorporate them in a solution as quantities that have 
uncertainties described by the full covariance matrix, while fixing them 
is equivalent to treating them as known exactly.

  The $\chi^2$ per the number of degrees of freedom statistics shows that 
group delay uncertainties computed by the fringe fitting algorithm,
$\sigma_a$, never match post-fit residuals. Using weights $w=1/\sigma_a$ 
results in very significant biases of estimates of \chalp{standard variations}{variances}
of adjusted parameters. To alleviate this problem, we used weights in 
the form $w=1/\sqrt{\sigma_a^2 + v^2}$, where $v$ is an extra variance 
computed for each experiment and for each baseline in such a way that 
the ratio of the weighted sum of postfit residuals to their mathematical 
expectation is close to unity following the approach first described
in \citet{r:ryan91}. The mathematical apparatus of this technique 
is presented in \citet{r:rfc1}. This additional variance $v$ turned 
out close to the weighted root mean square (wrms) of postfit 
residuals for many experiments.

\subsection{Geodetic data analysis of concurrent experiments}

  We ran a set of global solutions, including concurrent geodetic 
VLBI experiments. For each solution, the processed dataset was 
a concatenation of three lists: background experiments, target 
experiments, and reference experiments. The EOP epoch of each 
target experiment was determined as the mean epoch of the 
observations used in that experiment. For 1-hr experiments, only 
$E_3$ angles were estimated, but not rates, and therefore, the 
EOP epoch was implied to be the middle epoch of an observing
session. EOPs and their rates at the specific epoch were estimated 
from 24-hr reference experiments. That specific epoch was set to 
the EOP epoch of a corresponding target experiment. Similarly, when 
a target experiment had a duration of 24~hours, EOP rate and EOP 
at the explicitly specified epoch were estimated.

\begin{figure}
   \noindent
   \centerline{\includegraphics[width=0.33\textwidth]{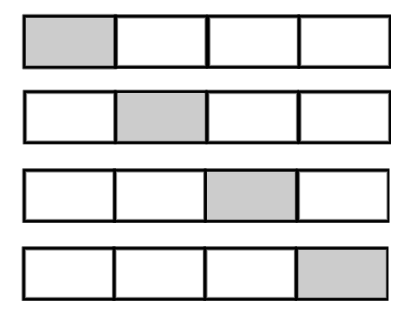}}
   \caption{Illustration of reprocessing a given experiment
            four times using non-overlapping ranges of data
            shown with gray color and downweighting all other 
            data.
           }
   \label{f:grey_boxes}
\end{figure}

  A given target experiment can be virtually shrunk by downweighting
all observations outside the specified range by a factor of 1000. 
We used the following technique: we processed a given experiment $N$
times using data from $N$ non-overlapping segments and downweighting
all other data by a factor of 1000. Figure~\ref{f:grey_boxes} 
illustrates this approach. This effectively transforms a long
experiment into a sequence of shorter experiments.

\section{Results}\label{s:res}

  Here we present our results of processing different sets of 
concurrent experiments from global solutions that used background,
target, and reference data.

\subsection{Results of s22 campaign}\label{s:res_s22}

   In program s22 twenty two consecutive 1-hr VLBI experiments at a single 
baseline Mg-Ws were observed in 22-hr sessions that are concurrent with 
multi-baseline 24-hr r1 experiments. Since there were several gaps, the 
total count of s22 experiments is 215. We estimated $E_3$ angles from 
s22 experiments and estimated EOP and their rates from concurrent r1
experiments on the middle epoch of s22 observing sessions.

   We computed biases and rms of differences in $E_3$ angles from each 
22-hr observing session and from all the data combined. Results are presented
in Table~\ref{t:s22}. Errors of both r1r4 and s22 campaigns contribute to
the rms of differences. We corrected these rms by subtracting in quadrature
the contribution to these differences from 24-hr programs that was set
to \chalp{0.53~nrad}{0.40~nrad} for all the experiments. We discuss how we 
obtained this contribution in subsection \ref{s:reg_24hr}. We call these 
statistics ``corrected''. \chalp{We consider the rms as a measure of EOP
accuracy. We report also biases, but we consider a bias and a scatter 
as two components of an error.}

   Even a  cursory analysis reveals a noticeable difference
between experiments ran in summer months May, June, July, August, September 
and winter months November, December, January, February, and April. 
According to the classical F-test, the ratio of \chalp{standard deviations}{rms}, 
1.85 is statistically significant at a $10^{-9}$ confidence level for 
a sample of this size. 

\begin{table}
     \caption{Estimates of biases and \chalp{rms}{std}of differences in $E_3$ 
              angles determined from 1-hr s22 experiments with respect 
              to concurrent multi-baseline 24-hr experiment. The \chalp{rms}{std}
              of the differences was corrected for the contribution
              of errors from 24-hr experiments.
             }
     \begin{tabular}{ll @{\hspace{0.5em}} rr r}
         \hline
         Session name  & Session code & bias  & \chalp{rms}{std}  & \# exp \\
                       &              & nrad  & nrad &        \\
         \hline                                        \\
         20220411\_c   & s22401 &  0.34 & 0.52 &  22   \\
         20231113\_a   & s22415 & -0.06 & 1.28 &  17   \\
         20231211\_a   & s22416 &  0.03 & 0.30 &  22   \\
         20240122\_a   & s22417 &  0.36 & 0.72 &  22   \\
         20240219\_b   & s22418 &  0.41 & 0.70 &  22   \\
         20240415\_b   & s22420 &  0.27 & 0.97 &  22   \\
         20240610\_b   & s22422 & -0.12 & 1.31 &  22   \\
         20240715\_b   & s22423 &  0.11 & 1.58 &  22   \\
         20240812\_b   & s22424 &  0.35 & 1.26 &  22   \\
         20240916\_b   & s22425 & -0.44 & 1.34 &  22   \\
         \hline                                        \\
         Winter        &        &  0.29 & 0.65 & 127   \\
         Summer        &        &  0.07 & 1.42 &  88   \\
         All           &        &  0.22 & 0.97 & 215   \\
         \hline                                        \\
     \end{tabular}
     \label{t:s22}
\end{table}

\begin{figure}
   \noindent
   \includegraphics[width=0.49\textwidth]{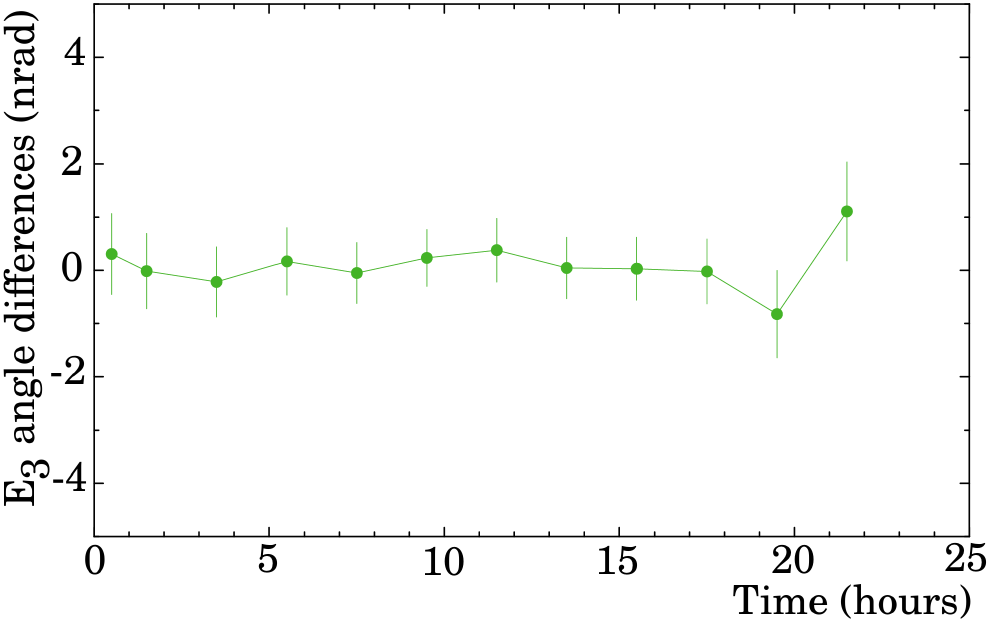}
   \hspace{0.01\textwidth}
   \includegraphics[width=0.49\textwidth]{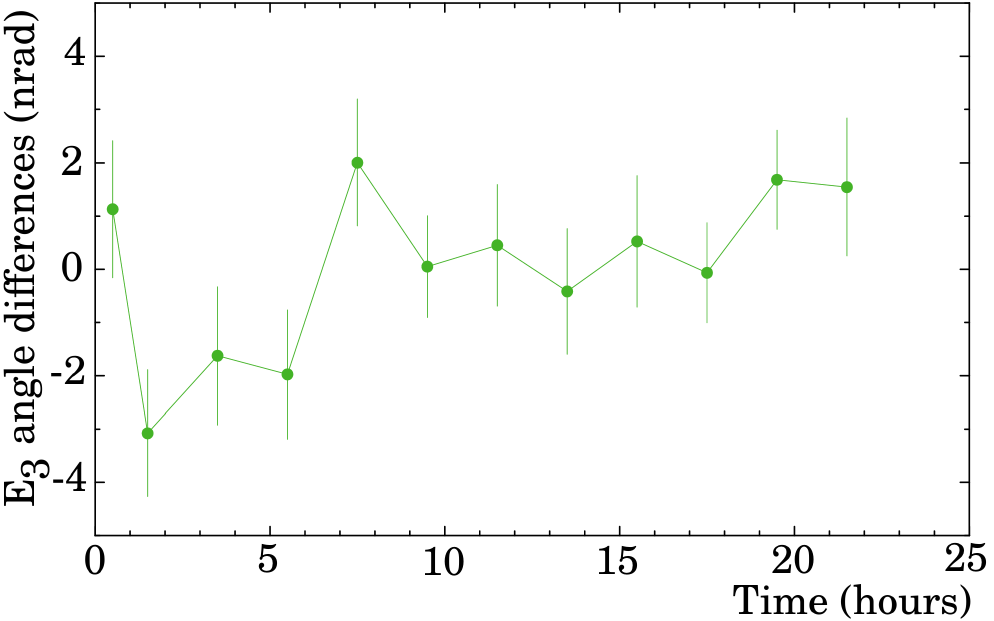}
   \caption{Differences in $E_3$ angle estimates between 22 single-baseline 
            experiments at MgWs baseline versus estimates from 
            concurrent ten-baseline 24-hr experiments. 
            {\it Left:}   experiment with  code s22416 on December 13, 2023. 
            {\it Right: } experiment with  code s22423 on July 15, 2024. 
           }
   \label{f:e3_mgws}
\end{figure}

  Figure~\ref{f:e3_mgws} illustrates the differences in $E_3$ angles from 
two observing sessions with the smallest and largest \chalp{standard deviations}{rms}.
The differences do not exhibit a noticeable systematic pattern. The biases
in $E_3$ angles do not exceed 40\% of a \chalp{standard deviation}{rms} for any 
observing session and they are at a level of 1/5 of the \chalp{standard 
deviation}{rms} of the whole dataset.

  We computed the rms of the differences for experiments scheduled 
according to strategy A and strategy B separately. We got 0.96 and 
0.97~nrad respectively, which is totally insignificant. The differences 
remained insignificant when winter only or summer only experiments 
were used.

\subsection{Results from regular 1-hr observing campaigns \remlp{at KkWz}}\label{s:reg_1hr}

\chalp{}{}
  The goal of 1-hr campaigns is to provide $E_3$ angle estimates with 
a short latency. For instance, the goal of IVS-INT-00 campaign is to have 
a latency of 36~hours, and this explains why these campaigns are called 
intensive. To meet these requirements, the experiment duration is reduced 
to one hour and the number of stations is reduced to two. Occasionally, 
a third and fourth station is added. A significant number of these 
experiments overlap with r1r4 24-hr observing sessions. We set EOP reference 
epochs for corresponding r1r4 experiments to the middle epoch of 1-hr 
experiments and derive $E_3$ angles at exactly the same epochs from 1-hr 
and 24-hr experiments. We computed statistics of the differences and 
applied an iterative procedure starting with the data point with the 
greatest $|x|/$rms ratio and consecutively eliminating such points and 
re-computing the rms till no data points with this ratio greater than 
3 remained. The number of outliers does not exceed 2\%. The statistics 
of the differences corrected for outliers, as well as the mean values 
of the corresponding formal errors are shown in Table~\ref{t:others_1hr}.

\begin{table}
   \caption{Differences in the $E_3$ angle estimates between regular
            single-baseline 1-hr experiments versus estimates from 
            the concurrent multi-baseline 24-hr experiments for the
            same epoch. The rms of the differences were corrected for
            the contribution of errors from 24-hr experiments.
            The last column provides the arithmetic mean of formal
            errors from our solutions.
           }
   \begin{tabular}{lr rrc}
       \hline                                             \\
       Baseline & \# exp & \ntab{c}{bias}  & \ntab{c}{rms}  & \ntab{c}{mean $\sigma_f$} \\
                &        & \ntab{c}{nrad}  & \ntab{c}{nrad} & nrad                      \\
       \hline                                             \\
       K2Ws &            409 &  0.29  & \chalp{1.35}{1.40} & 0.59 \\
       MgWs &             29 & -0.17  & \chalp{1.30}{1.33} & 0.62 \\ 
       MgWs${}^\star$ &  215 &  0.17  & \chalp{1.04}{0.97} & 0.66 \\ 
       TsWz &            107 &  0.21  & \chalp{1.45}{1.54} & 0.99 \\ 
       IsWz &             65 &  0.74  & \chalp{1.43}{1.53} & 0.86 \\ 
       MkWz &            115 &  0.04  & \chalp{1.21}{1.32} & 0.72 \\ 
       HnMk &             58 &  0.75  & \chalp{1.60}{1.68} & 1.36 \\ 
       KkWz &           2905 & -0.22  & \chalp{1.50}{1.59} & 1.26 \\ 
       \hline                                             \\
   \end{tabular}
   {\raggedright 
                 MgWs${}^\star$ denotes estimates from s22 program, while 
                 MgWs\phantom{${}^\star$} denotes estimates from other programs.
   }         
   \label{t:others_1hr}
\end{table}

\begin{figure}
   \noindent
   \includegraphics[width=0.50\textwidth]{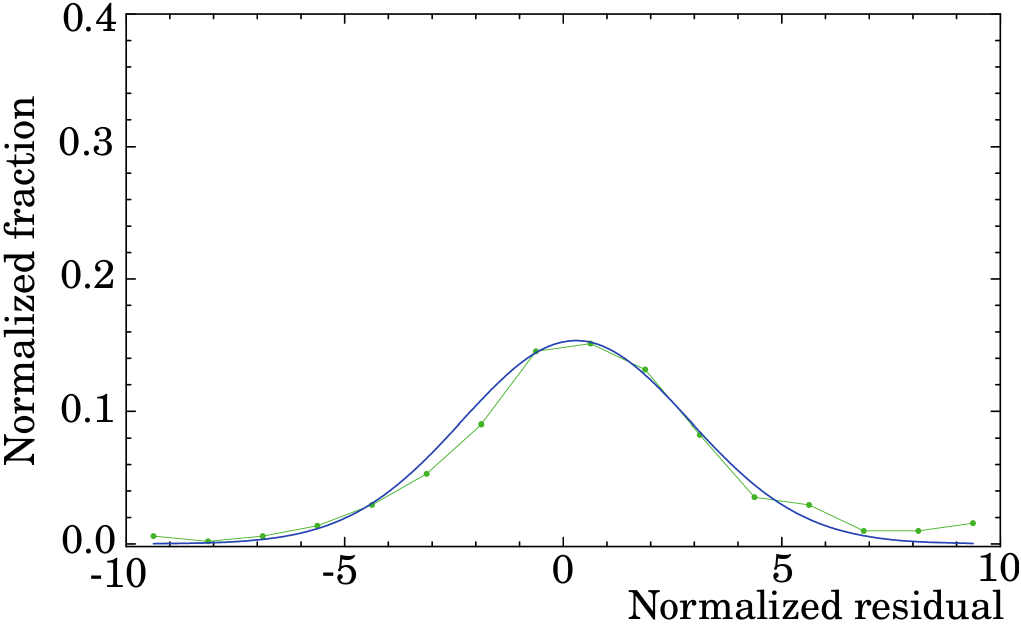}
   \hspace{0.01\textwidth}
   \includegraphics[width=0.48\textwidth]{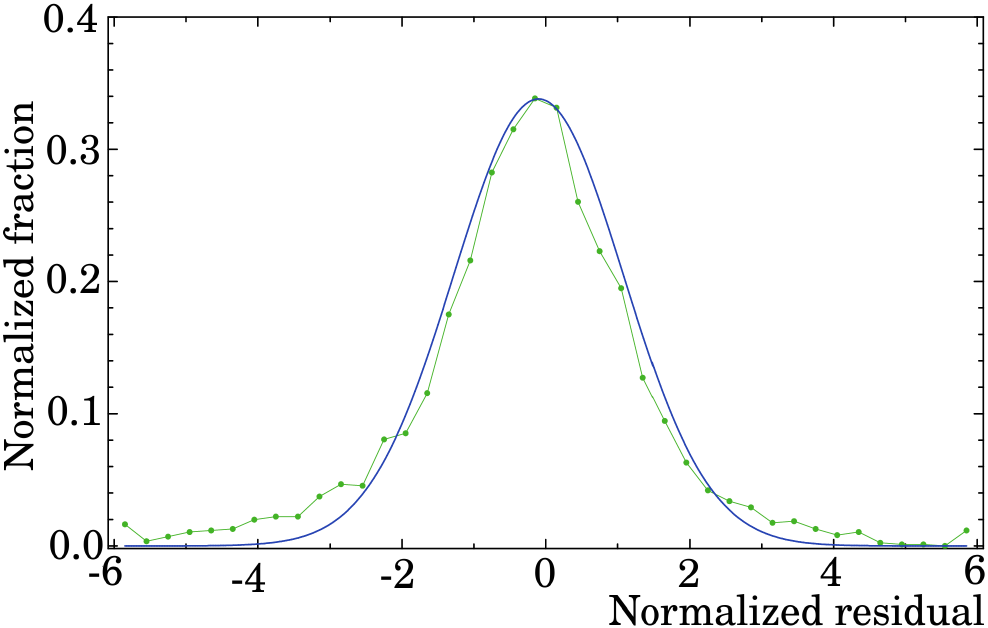}
   \caption{The normalized distributions of the $E_3$ angle differences
            derived from analysis of experiments at K2Ws (left) and KkWz 
            baselines (right) divided by formal uncertainties. The blue 
            smoothed lines show Gaussian distributions with the second 
            moments 2.60 and 1.18 respectively.
           }
   \label{f:e3_nrml_err}
\end{figure}

   We computed the distribution of residuals divided by scaled 
uncertainties and show the plot for experiments at baseline K2Ws and 
KkWz in Figure~\ref{f:e3_nrml_err}. For comparison, we also plot 
the Gaussian distributions with the first moments equal to the bias and 
with fitted second moments. We see that the distribution of normalized 
differences of $E_3$ angles determined from observations at KkWz 
baselines has a positive excess kurtosis, 0.84, i.e. that distribution
has heavier tails than the normal  distribution in the regions of 
normalized residuals that are by a modulo above 2.5. This is a common 
feature of empirical distributions. The fitted second moment, 2.60 for 
K2Ws and 1.18 of KkWz baseline is within 5--10\% of the ratio of the 
rms to the mean formal uncertainty. We should note that baselines K2Ws 
and KkWz are geometrically almost identical. The main difference is the 
recorded bandwidth and therefore, the reported group delay uncertainties. 

\begin{figure}
   \centerline{\includegraphics[width=0.50\textwidth]{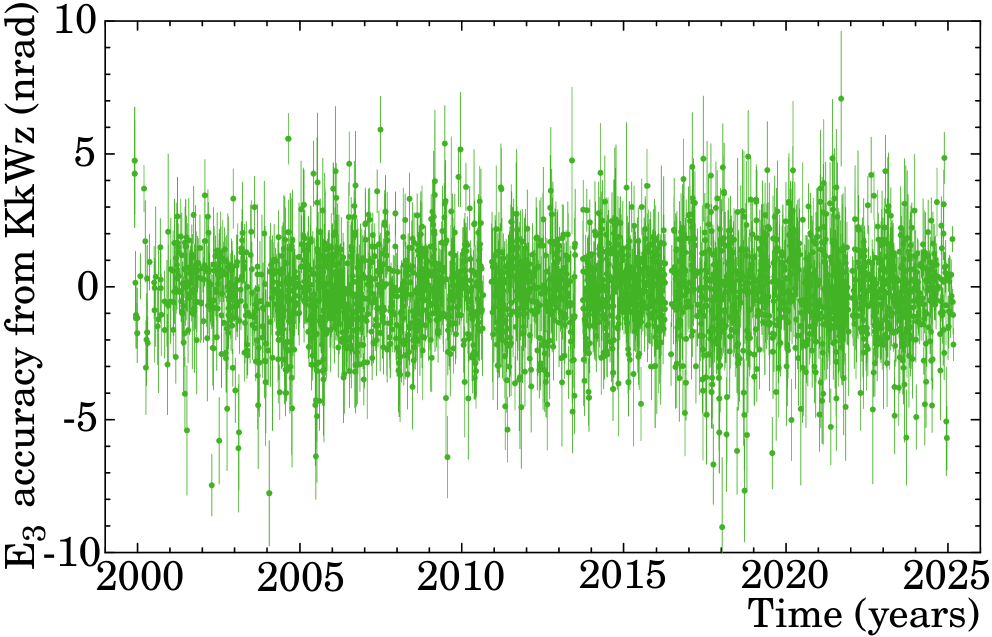}}
   \caption{Differences in the $E_3$ angle estimates derived from 
            processing data at single-baseline 1-hr experiments at KkWz 
            baseline versus concurrent multi-baseline 24-hr r1 experiments 
            as a function of time.
           }
   \label{f:e3_kkwz}
\end{figure}

   We show in Figure~\ref{f:e3_kkwz} the differences in $E_3$ angles from 
the \chalp{INT-INT-1}{IVS-INT-1} campaign that has been running for over 25~years. We do not 
see a significant systematic pattern in the plot of residuals.
Then we computed the \chalp{rms}{std} over six summer months May, June, July, August,
September, October and over six winter moths. The rms of differences
are \chalp{1.57 and 1.39~nrad}{1.66 and 1.49~nrad} respectively. Since
the number of winter and summer
concurrent experiments is over 1400, the F-test shows that this difference
is significant at a $10^{-5}$ level. Winter IVS-INT-1 experiments are more 
accurate than summer experiments. Summer experiments on average add 
in quadrature 0.73~nrad variance.

  We performed a similar summer/winter test for VGOS-INT-A campaign at 
baseline K2Ws. Both ends of baseline K2Ws and KKWz are located within less 
than 100~meters of each other. The corrected rms of the differences are 
\chalp{1.11~nrad for winter and 1.42~nrad}{1.23~nrad for winter and 1.52~nrad}
for summer. According to F-test these differences are significant at
a $6 \cdot 10^{-4}$ level. Summer experiments on average add in
quadrature 0.89~nrad variance \addlp{with respect
to winter experiments.}

\subsection{Results from regular concurrent 24-hr VLBI campaigns}\label{s:reg_24hr}

   We have processed data from 24-hr concurrent VLBI campaigns in 
a similar way. We estimated the EOP and their rates on epochs that are 
the arithmetic mean of the middle epochs of each individual experiment. 
We computed the mean value of the differences (bias) and their rms. 
Since a~priori we do not have knowledge that EOP estimates from one of 
the campaigns vo versus r1r4 or ca versus na are substantially 
different, we assume they are equally accurate and uncorrelated. Since 
observing stations of concurrent networks are typically separated by 
hundreds of kilometers and more, we consider that atmospheric path delay
is expected to be uncorrelated at these distances to a great extent.
Therefore, we divide the rms by $\sqrt{2}$ and report in 
Table~\ref{t:conc_24hr} the scaled rms which we view as a measure 
of the average accuracy of concurrent vo/r1r4 and ca/na campaigns.
  
\begin{table}
     \caption{Accuracy of EOP determination from comparison of 24-hr
              concurrent observing sessions under ca/na and 
              vo/r1r4 programs. The rms of differences were divided by 
              $\sqrt{2}$.
             }
     \begin{tabular}{lc rl lr}
        \hline \\
        Campaign & Comp & bias & rms & units & \# points \\
        \hline \\
        ca/na & $ E_1       $ &  0.30 & 0.57 & \addlp{$ 10^{-9}  $} rad   &  82 \\
        ca/na & $ E_2       $ & -0.06 & 0.83 & \addlp{$ 10^{-9}  $} rad   &  82 \\
        ca/na & $ E_3       $ &  0.03 & 0.59 & \addlp{$ 10^{-9}  $} rad   &  82 \\
        ca/na & $ \dot{E}_1 $ &  0.59 & 2.26 & \addlp{$ 10^{-14} $} rad/s &  82 \\
        ca/na & $ \dot{E}_2 $ & -0.30 & 2.23 & $ 10^{-14} $ rad/s &  81 \\
        ca/na & $ \dot{E}_3 $ & -0.43 & 1.45 & $ 10^{-14} $ rad/s &  82 \\ 
        \vspace{-1.50ex} \\
        vo/r1r4     & $ E_1       $ &  \chalp{0.11}{0.57} & \chalp{0.51}{0.57} & \addlp{$ 10^{-9}  $} rad   & \chalp{154}{150} \\
        vo/r1r4     & $ E_2       $ &  \chalp{0.42}{0.63} & 0.56 & \addlp{$ 10^{-9}  $} rad   & \chalp{156}{150} \\
        vo/r1r4     & $ E_3       $ &  \chalp{0.13}{0.40} & 0.40 & \addlp{$ 10^{-9}  $} rad   & \chalp{157}{150} \\
        vo/r1r4     & $ \dot{E}_1 $ &  0.00 & \chalp{1.66}{1.63} & $ 10^{-14} $ rad/s & \chalp{155}{150} \\
        vo/r1r4     & $ \dot{E}_2 $ &  0.01 & \chalp{1.47}{1.51} & $ 10^{-14} $ rad/s & \chalp{156}{150}\\
        vo/r1r4     & $ \dot{E}_3 $ &  0.00 & \chalp{1.12}{1.02} & $ 10^{-14} $ rad/s & \chalp{157}{150} \\
        \vspace{-1.50ex} \\
        \hline 
     \end{tabular}
     \label{t:conc_24hr}
\end{table}

  We see the biases are less than $0.5\sigma$, i.e. insignificant. 
It is worth noting that although the rms of differences of 
$E_1$ and $E_2$ from vo/r1r4 campaigns is smaller than from ca/na 
campaigns that ran 25~years earlier, this difference is not statistically 
significant even at a 0.20 level. The differences in $E_3$ accuracies
are significant at a 0.002 level and in $\dot{E}_3$ are significant 
at a 0.005 level: vo/r1r4 results are more accurate. The 
differences in rms of $\dot{E}_1$ and $\dot{E}_2$ are significant 
at $10^{-5}$.

  $E_1$, $E_2$, and $\dot{E}_3$ are regularly estimated during analysis of
GNSS data by a number of groups. This prompted us to use the three corner
hat method \citep{r:tch} for evaluation of uncertainties of these 
parameters from GNSS, vo, and r1r4 campaigns. Indeed, when three time 
series are uncorrelated, we can derive the variances of each series from 
the variances of their differences. We took daily time series of polar 
motion and lengths of day from the IGS final series, converted them to 
$E_1$, $E_2$, and $\dot{E}_3$, interpolated to the epoch of observations, 
computed \chalp{the rms}{the bias and rms} of the differences, and then derived standard deviations 
of each three series. Results are presented in Table~\ref{t:three_corner}. 
We are aware that the three corner hat method is susceptible to the 
presence of weak correlation and systematic errors. We computed 
a smoothing spline at 9~equally distributed knots to fit the difference 
between the series and removed a small systematic pattern, but this did 
not change the results noticeably.

  We see that the accuracy of polar motion and $\dot{E}_3$ from GNSS has
improved by a factor of 2.5 over 25~years since 1997--2000.5. Our accuracy 
estimate, 0.20~nrad for $E_1$ and $E_2$, is 50\% higher than the estimate 
of \citet{r:ray17} (0.12--0.15~nrad). The EOP accuracy from VLBI also 
improved, but the improvement is not that strong: 28 to 68\%. 
The results of our comparison  shows that accuracy of $E_1$ and $E_2$ 
from r1r4 sessions is better than both accuracy from old ca/na campaigns 
and from vo campaign. According to the F-test, this accuracy difference 
is significant at a level of 0.002. The difference in uncertainties of 
$\dot{E}_3$ from these campaigns is significant at the 0.01 level. 
Our tests confirm that accuracy of $E_1$ and $E_2$ from GNSS was on par 
with VLBI in late 1990s, but it improved by a factor of 2--3 by the 2020s.

\begin{table}
     \caption{Upper table: three corner hat analysis of rms of the 
              estimates of $E_1$, $E_2$, $\dot{E}_3$ from GNSS, ca,
              and na VLBI programs over 1997.0--2000.5. 
              Lower table: the same analysis for GNSS, vo, 
              and r1r4 VLBI programs over 2019.0--2024.0. 
              Column~2: rms of differences between GNSS results (IGS Final) 
                        versus results from VLBI ca or vo experiments.
              Column~3: rms of differences between GNSS results 
                        versus results from VLBI na or r1r4 experiments.
              Column~4: rms of differences between VLBI results from ca versus
                        na experiments or from vo versus r1r4 experiments.
              Columns 5--7: rms of estimates of $E_1$, $E_2$, $\dot{E}_3$ 
                            derived by the three hat corner algorithm.
              Columns 5: rms of GNSS results.
              Columns 6: rms of results from VLBI ca or vo experiments.
              Columns 7: rms of results from VLBI na or r1r4 experiments.
              Columns 8: unit. 
             }
     \begin{tabular}{l lll @{\qquad} lll l}
           \hline \\
            \ntab{c}{(1)} & \ntab{c}{(2)} & \ntab{c}{(3)} & \ntab{c}{(4)} &
            \ntab{c}{(5)} & \ntab{c}{(6)} & \ntab{c}{(7)} & \ntab{c}{(8)} \\
            Comp  & $\sigma_{\rm gnss,ca}$ & $\sigma_{\rm gnss,na}$ & $\sigma_{\rm ca,na}$ & 
                    $\sigma_{\rm gnss}$    & $\sigma_{\rm ca}$      & $\sigma_{\rm na}$ &
                  unit \\
           \hline \\
                $E_1$ & 0.71  &  0.65  & 0.66 &  0.50 &  0.51 &  0.42 & $10^{-9}\:\;$ rad    \\
                $E_2$ & 0.63  &  0.67  & 0.73 &  0.40 &  0.49 &  0.54 & $10^{-9}\:\;$ rad    \\
          $\dot{E}_3$ & 3.26  &  2.93  & 2.10 &  2.72 &  1.80 &  1.08 & $10^{-14}$ rad/s     \\
           \hline \\
            Comp  & $\sigma_{\rm gnss,vo}$ & $\sigma_{\rm gnss,r1r4}$ & $\sigma_{\rm vo,r1r4}$ & 
                    $\sigma_{\rm gnss}$    & $\sigma_{\rm vo}$      & $\sigma_{\rm r1r4}$ &
                  unit \\
           \hline \\
                $E_1$ & 0.52  &  0.38  & 0.60 &  0.16 &  0.46 &  0.34 & $10^{-9}\:\;$ rad    \\
                $E_2$ & 0.62  &  0.40  & 0.65 &  0.24 &  0.57 &  0.32 & $10^{-9}\:\;$ rad    \\
          $\dot{E}_3$ & 1.45  &  1.34  & 1.31 &  1.17 &  1.04 &  0.84 & $10^{-14}$ rad/s     \\
           \hline
           \hline
     \end{tabular}
     \label{t:three_corner}
\end{table}

\subsection{Results from 1-hr blocks of concurrent 24-hr VLBI campaigns}\label{s:reg_24_1hr}

   In subsection \ref{s:res_s22} we explained how we compared a dedicated 
campaign s22 that contained 22 1-hr schedules in one observing session. 
These 1-hr schedules were designed independently. Here we expand this 
approach to 24~hours experiments. These experiments were scheduled as 
single 24-hr blocks and were not designed to be processed as 1-hr 
separate blocks. Nevertheless, we consider that it would be instructive 
to analyze concurrent experiments that way.

   We reprocessed two concurrent campaigns, ca/na and vo/r1r4. 
Results are shown in Table~\ref{t:conc_24hr_1hr}. It is worth
comparing these rms of the differences against Table~\ref{t:conc_24hr} 
with the results of processing whole 24-hr data chunks. As expected,
shrinking the dataset by a factor of 24 caused an increase in 
rms of EOP estimates, in particular, a factor of 3.7 for ca/na and 2.8 
for vo/r1r4 for $E_3$. But this reduction of rms is short of a naive
$\sqrt{24} \approx 4.9$ factor.

\begin{table}
     \caption{Accuracy of EOP determination from comparison of 24-hr
              observing sessions under programs vo/r1r4 and ca/na
              from 1-hr segments. The rms of differences were divided 
              by $\sqrt{2}$.
             }
     \begin{tabular}{lrr rrr}
        \hline \\
        Campaign & duration & \# seg &  \ntab{c}{$E_1$} & \ntab{c}{$E_2$} & \ntab{c}{$E_3$} \\
                 & hour     &        &            nrad  &           nrad  &           nrad  \\
        \hline                                                                              \\
        ca/na    &  1       &   1763 &            3.53  &           3.90  &           2.16  \\
        vo/r1r4  &  1       &   3348 &            1.61  &           1.81  &           1.23  \\
        \hline                                                                              \\
     \end{tabular}
     \label{t:conc_24hr_1hr}
\end{table}

\begin{figure}
   \noindent
   \includegraphics[width=0.48\textwidth]{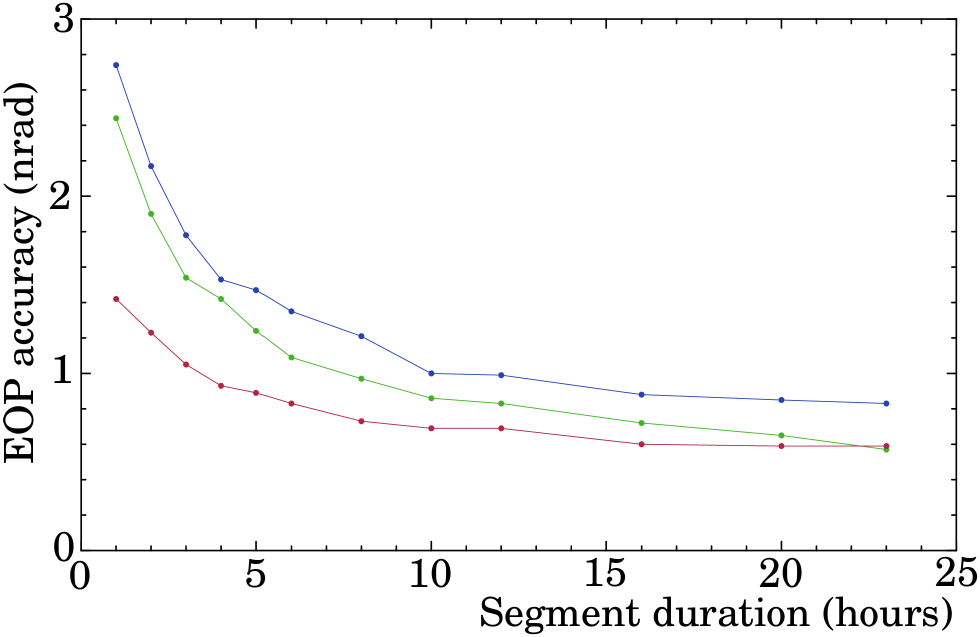}
   \hspace{0.01\textwidth}
   \includegraphics[width=0.48\textwidth]{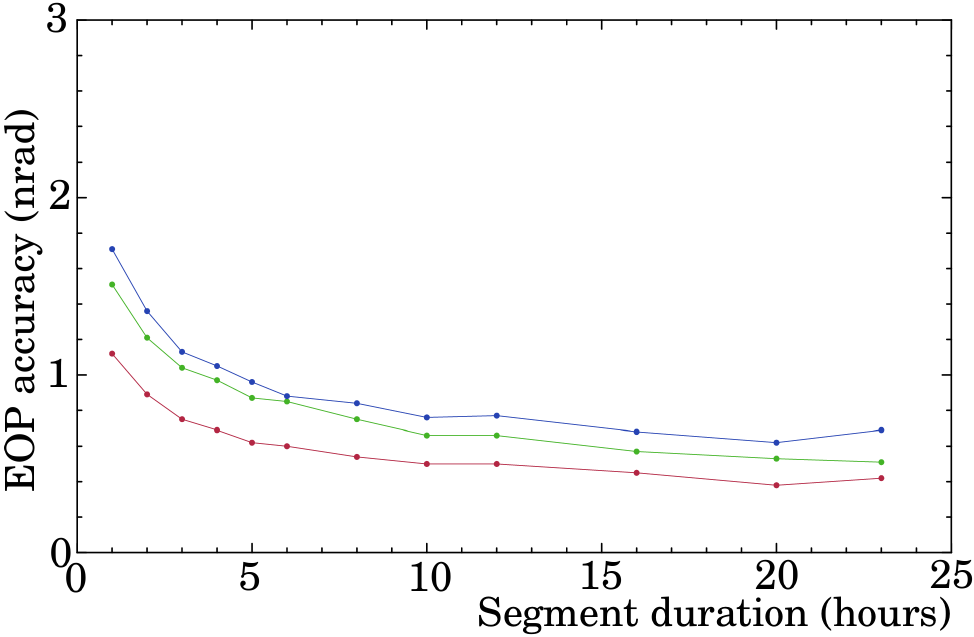}
   \caption{rms of errors in $E_1$ (middle green line), $E_2$
            (upper blue line), $E_3$ (red bottom line) as a function 
            of segment duration. The left plot shows results for ca/na
            campaigns. The right plot shows results for vo/r1r4 campaigns.
           }
   \label{f:e1_e2_e3_span}
\end{figure}

  We found this phenomenon deserves further investigation. We ran a 
battery of solutions and varied segment duration from 1 hour to 23~hours 
with a step of 0.5 hour in a range of 1~to 6~hours and with a step of 
1~hour for segments longer than 6~hours. The dependence of the rms of 
differences in EOP estimates is a monotonously decaying function of 
the segment length. The EOP accuracy as a function of the segment 
length for ca/na and vo/r1r4 programs is shown 
in Figure~\ref{f:e1_e2_e3_span}.

\begin{figure}
   \noindent
   \includegraphics[width=0.48\textwidth]{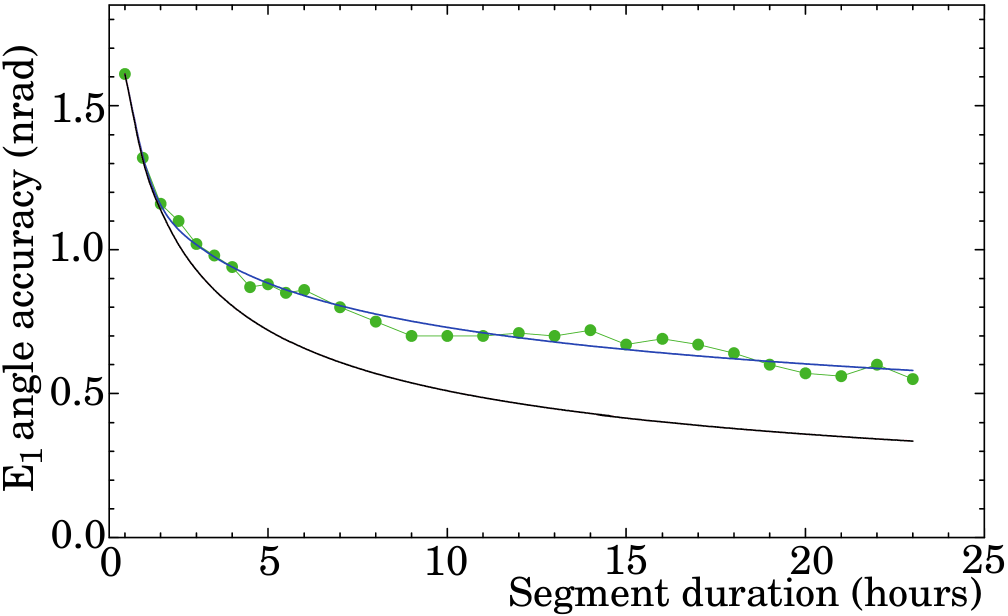}
   \hspace{0.01\textwidth}
   \includegraphics[width=0.48\textwidth]{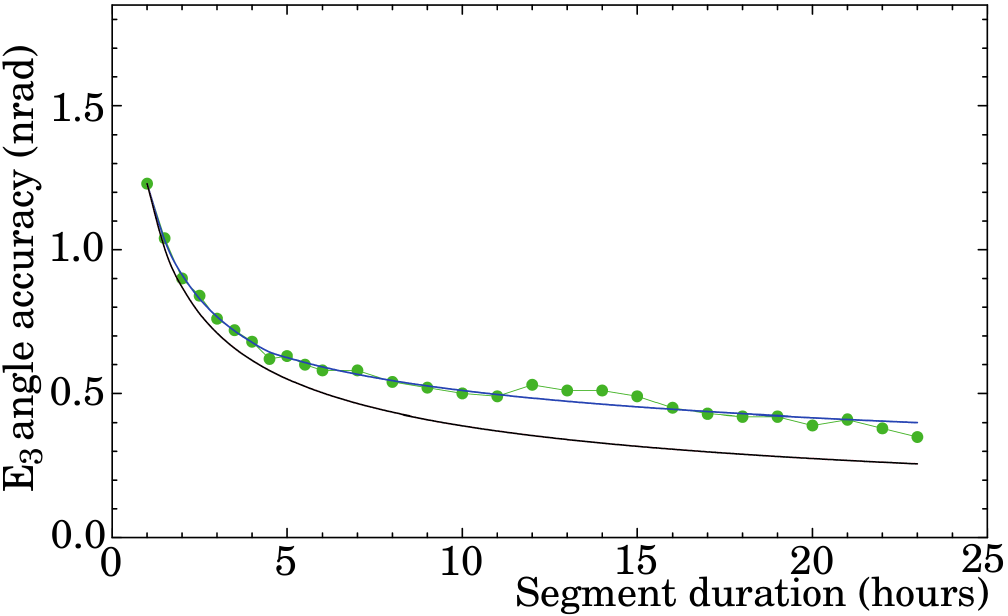}
   \caption{The rms of errors in estimates of $E_1$ (left) and $E_3$ (right) 
            angles as a function of experiment duration from vo/r1 
            comparison (green dots). The middle blue line shows the best fit 
            of the broken power laws, while the bottom black line shows 
            the power law with $\beta=-0.50$, representing the expectation 
            based on uncorrelated observations. 
           }
   \label{f:e3_v_span}
\end{figure}

  We tried to approximate this dependence with a smooth curve. Since we do 
not know a~priori its functional dependence, adhering the Occam's principle, 
we were looking for the simplest form, such as an exponential decay or 
a power law. Neither of these simple functions fit the data. We resorted to 
the broken power law that is commonly used in astrophysics in the form of
\beq
    f(t) = \left\{ \begin{array}{l}
                       t^{\beta_1} \enskip {\rm if} \enskip t >    b  \vex \\
                       t^{\beta_2} \enskip {\rm if} \enskip t \leq b
                   \end{array}
           \right.,
\eeq{e:e5}

   where $b$ is a breaking point. We found numerically $b$, $\beta_1$, and 
$\beta_2$ that minimize the sum of squares of residuals on a grid using 
a brute force approach. The results for $E_1$ and $E_3$ angles are shown in 
Figure~\ref{f:e3_v_span}. 

   We see that for segments shorter than 2--3 hours, the rms 
of EOP differences decreases quite fast with an increase in the amount 
of data in a rate that is comparable with a square root of the number 
of data samples. For longer segments, the rms of EOP differences decreases 
much slower than the square root power law. It follows from the theorem 
of Gauss-Markov that if the noise is uncorrelated, then the variance of 
least square estimates for a given set of estimated parameters is reduced 
as square root of the number of observations. Therefore, we interpret 
Figure~\ref{f:e3_v_span} as a direct evidence of a violation of the 
assumption that the noise is uncorrelated.

  In order to confirm this hypothesis, we ran another set of solutions
and added uncorrelated, Gaussian, zero-mean noise produced by the random
noise generator with specified second moments. When we add a noise with
known characteristics, we dilute the properties of the intrinsic noise.
We expect that, according to the Gauss-Markov theorem, the dependence
of the EOP accuracy as a function of segment length should asymptotically
converge to the power law -0.5 when the amount of the injected uncorrelated
noise becomes dominant. Results are presented in Table~\ref{t:eop_bl_model}.
We see that indeed, adding uncorrelated noise made estimates of $\beta_2$
closer to -0.5. This pattern is seen more clearly for $E_3$ angle than
for $E_1$ and $E_2$. Parameter $\beta_2$ converges faster for 
ca/na experiments compared to vo/r1r4. 

\begin{table}
     \caption{The estimates of the model of scaling EOP errors
              with the observing session duration with a broken
              power laws with powers $\beta_1$ at spans less 
              than the breaking point $b$ and $\beta_2$ after 
              that. Computation is done for two concurrent 
              campaigns vo/r1r4 and ca/na in four modes: 
              when original data are used and when 
              uncorrelated white noise with the second moment
              20, 50, and 100~ps was added.
             }
     \begin{tabular}{l rcl @{\quad\quad} l rcl}
        \hline                                        \\
        Component  & $\beta_1$ & $b$ & $\beta_2$ &  
        Component  & $\beta_1$ & $b$ & $\beta_2$      \\
        \hline                                                                              \vspace{-0.5ex} \\
        $E_1$ vo/r1r4         & -0.478 & 2.16 & -0.276  & $E_1$ ca/na         & -0.895 & 1.70 & -0.510  \\
        $E_2$ vo/r1r4         & -0.460 & 2.48 & -0.303  & $E_2$ ca/na         & -0.765 & 2.13 & -0.444  \\
        $E_3$ vo/r1r4         & -0.430 & 4.50 & -0.292  & $E_3$ ca/na         & -0.693 & 2.87 & -0.331  \vspace{0.75ex} \\
        $E_1$ vo/r1r4 + 20ps  & -0.570 & 1.74 & -0.308  & $E_1$ ca/na + 20ps  & -0.792 & 2.05 & -0.517  \\
        $E_2$ vo/r1r4 + 20ps  & -0.481 & 2.55 & -0.330  & $E_2$ ca/na + 20ps  & -0.744 & 3.30 & -0.427  \\
        $E_3$ vo/r1r4 + 20ps  & -0.526 & 2.20 & -0.345  & $E_3$ ca/na + 20ps  & -0.692 & 2.68 & -0.392  \vspace{0.75ex} \\
        $E_1$ vo/r1r4 + 50ps  & -0.506 & 3.19 & -0.311  & $E_1$ ca/na + 50ps  & -0.782 & 3.00 & -0.531  \\
        $E_2$ vo/r1r4 + 50ps  & -0.502 & 2.61 & -0.344  & $E_2$ ca/na + 50ps  & -0.734 & 3.50 & -0.534  \\
        $E_3$ vo/r1r4 + 50ps  & -0.575 & 2.85 & -0.374  & $E_3$ ca/na + 50ps  & -0.709 & 3.07 & -0.505  \vspace{0.75ex} \\
        $E_1$ vo/r1r4 + 100ps & -0.563 & 3.81 & -0.341  & $E_1$ ca/na + 100ps & -0.763 & 4.21 & -0.539  \\
        $E_2$ vo/r1r4 + 100ps & -0.821 & 1.71 & -0.445  & $E_2$ ca/na + 100ps & -0.806 & 2.06 & -0.606  \\
        $E_3$ vo/r1r4 + 100ps & -0.668 & 1.87 & -0.488  & $E_3$ ca/na + 100ps & -0.771 & 2.19 & -0.628  \vspace{1.50ex} \\
        \hline                                                  \\
     \end{tabular}
     \label{t:eop_bl_model}
\end{table}

\subsection{Results from 0.5-hr blocks of 1-hr VLBI campaigns}\label{s:0.5hr}

  In a similar way, as we split 24-hr observing sessions into 1-hr
blocks, we split the 1-hr observing sessions at K2Ws baseline
into two 0.5-hr blocks. We selected that campaign because, on average,
there are 60 observations in each 1-hr experiment, and the experiments
were designed to observe low and high elevation sources every 
10--30~minutes. We computed $E_3$ angles using only the first 30~minutes of 
data and the last 30~minutes of data. Then we computed $E_3$ angles from 
concurrent 24-hr observations, formed the differences, and computed the 
rms. The rms corrected for errors in $E_3$ from 24-hr observing sessions 
over 346 epochs is 1.42~nrad. In comparison, when the full 1-hr blocks in 
KsWs were used, the resulting corrected $E_3$ rms is 1.30~nrad. An increase 
in the number of observations used in a solution by a factor of 2 caused 
a reduction of errors by 9\%. However, according to the F-test, this 
difference is statistically insignificant even at a 10\% level.

  We also computed statistics of $E_3$ angles for the same experiments from
two 0.5-hr blocks from the same experiment against each other by forming
differences $E_3(t_2) - E_3(t_1) - \dot{E}_3(t1) \; (t_2 - t_1)$. We took
$\dot{E}_3(t1)$ from results of processing 24-hr experiments. We found that
the rms of the differences is 1.47~nrad. 

\section{Discussion}\label{s:disc}

  Here we provide interpretation of our results.

\subsection{Outcomes of different scheduling strategies}\label{s:sche}

   Within the s22 experiments, we explicitly tested an observing strategy 
that we expected would provide a more precise evaluation of the 
atmospheric path delay from the observations themselves, because we 
included observations of sources at high and low elevation angles in 
a rapid sequence over short periods of time. We generated simulated 
datasets from schedules and added the noise based on the mathematical 
model of atmospheric turbulence and its code developed 
by \citet{r:pany11}. We ran a battery of least square solutions of the 
simulated datasets with a diagonal weight matrix, generated time series 
of simulated $E_3$ estimates and computed their rms: 1.24 and 1.21~nrad 
for scheduling strategies A and B respectively.

   Our analysis of scheduling blocks designed with an advanced
strategy revealed no significant improvement compared to results from
processing scheduling blocks designed according to the traditional
strategy. Does it mean that estimation of the atmospheric path delay 
plays no role? To answer these question, we ran two additional solutions
and processed 215 s22 sessions in two modes: 1) we estimated a single 
atmospheric path delay for 1-hr blocks and 2) we did not estimate
zenith path delay. Results of these two auxiliary solutions, as well
as the reference solution, are presented in Table~\ref{t:aux_s22}.

\begin{table}
   \caption{Impact of estimation of the atmospheric path delay in zenith
            direction on $E_3$ estimates from s22 experiments.
            The first row corresponds to the main solution with 
            estimation of path delay in zenith direction with a linear
            spline with segment lengths of 10 minutes. The second row 
            corresponds to the case when one parameter of atmospheric path 
            delay in zenith per a 1-hr block was estimated. And the last 
            row corresponds to a case when atmospheric path delay was
            not estimated. The third column labeled accuracy is the rms 
            of $E_3$ angle estimates from concurrent VLBI experiments.
           }
   \begin{tabular}{l c c c c}
        \hline
          Solution         & bias  & accuracy  & formal error & postfit rms \\
                           & nrad  & nrad      & nrad         & ps          \\
        \hline
          10 min segments  & 0.22  & \chalp{1.04}{0.97}  &        0.74  &    25.61    \\
          60 min segments  & 0.19  & \chalp{1.04}{0.97}  &        0.69  &    29.89    \\
          no estimation    & 3.35  & \chalp{1.46}{1.49}  &        0.66  &    56.31    \\
        \hline
   \end{tabular}
   \label{t:aux_s22}
\end{table}

  We see estimation of the atmospheric path delay in zenith direction
does reduce postfit statistics, eliminates biases, and improves the 
accuracy, but degrades formal errors. However, estimating atmospheric
path delay every 10 minutes in the reference solution makes 
no differences with respect to estimating it every 60 seconds.

  We interpret these results as a manifestation of a certain limit
of our ability to estimate atmospheric path delay from observations
themselves. Atmospheric turbulence does not vanish at scales less than
60~minutes, but modeling slant atmospheric path delay with the zenith 
path delay scaled by the atmospheric mapping function shows it does.
To explain this contradiction, we suggest that modeling residual 
atmospheric path delay as a product of the zenith path delay and
the mapping function is not adequate at scales shorter than one hour.

  It is worth noting that if the minimum formal uncertainty were chosen 
as the criterion for selection of the optimal estimation strategy, one 
could select an approach of not estimating atmospheric path delay in 
zenith direction, which in fact, provides the highest bias and highest 
rms of differences between $E_3$ estimates from concurrent VLBI 
experiments.

\subsection{Seasonal accuracy variability}

\begin{figure}
   \centerline{\includegraphics[width=0.61\textwidth]{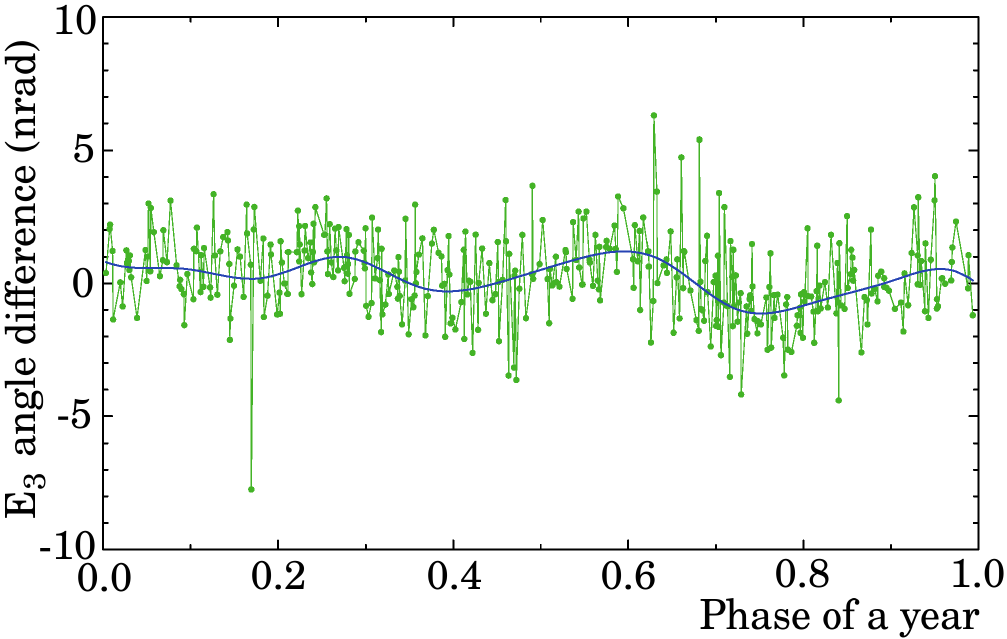}}
   \caption{Differences in the $E_3$ angle estimates between single-baseline 
            experiment at K2Ws baseline versus concurrent multi-baseline 
            24-hr r1 experiments as a function of a fractional part of 
            a year.
           }
   \label{f:k2_ws_phs}
\end{figure}

   Our analysis of 1-hr experiments at MgWz and KkWz reveals a strong
seasonality in the accuracy of $E_3$ angles estimates: the accuracy is 
worse in summer than in winter by the root sum squared 1.2 and 0.7~nrad
for these two baselines, and this difference is highly statistically
significant. However, the overall picture is not that clear.
We show in Figure~\ref{f:k2_ws_phs} the seasonal pattern by plotting
the difference in $E_3$ angle estimates from K2Ws baseline with
respect to the multi-baseline 24-hr experiments as a function of 
a fractional part of the year defined as the ratio of the day of a year
to 365. The fractional part of a year 0.0 corresponds to the 1st January.
The pattern shows several shallow maxima and minima.

\begin{figure}
   \includegraphics[width=0.49\textwidth]{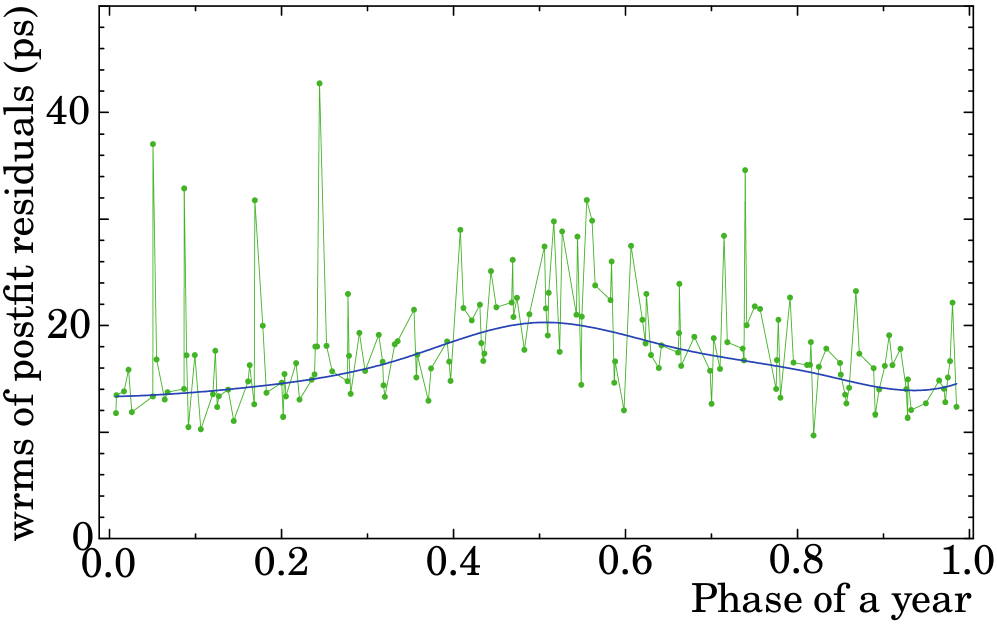}
   \hspace{0.01\textwidth}
   \includegraphics[width=0.49\textwidth]{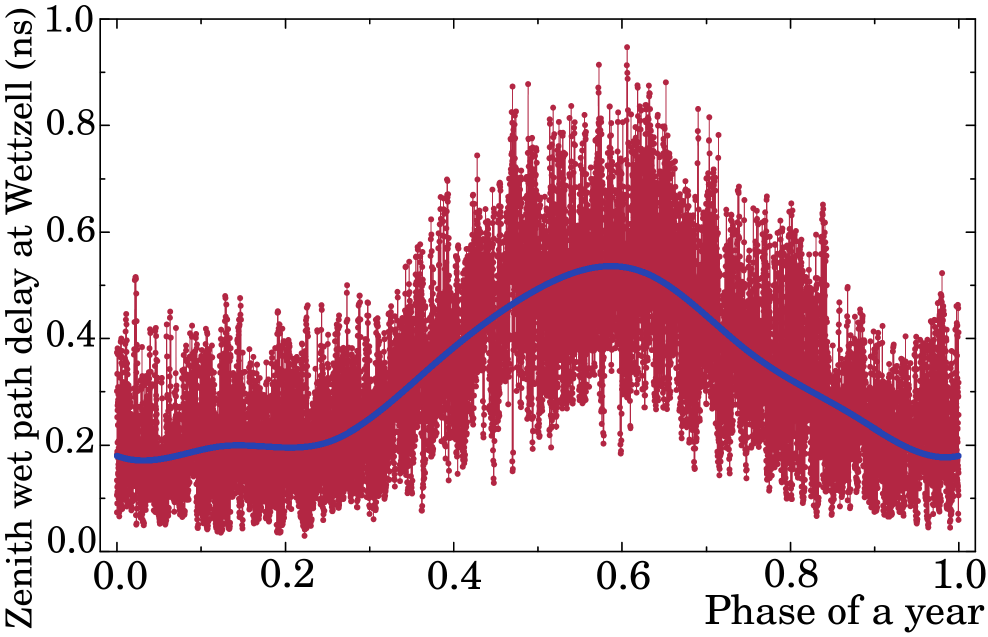}
   \caption{{\it Left: } the rms of postfit residuals of 24-hr vo experiments as 
            a function of a fractional part of a year.
            {\it Right: } zenith wet path delay at station Wz as 
            a function of a fractional part of a year.
           }
   \label{f:vo_rms_postfit_phs}
\end{figure}

  We noticed in 1990s that the scatter of the postfit residuals from 
geodetic observations at very sensitive Very Long Baseline Array (VLBA) 
antennas is noticeably greater in summer than in winter. 
The left plot in Figure~\ref{f:vo_rms_postfit_phs} demonstrates this pattern 
for modern quad-band vo VLBI experiments. The rms of postfit residuals in July
is a factor of 1.6 greater than in January. For comparison, we show a plot of 
zenith path delay at Wz stations also as a function of the fractional part 
of a year.

  Putting together these two findings, we conclude that the dominant 
error source in EOP estimates is due to the residual contribution of 
the atmospheric path delay that was not captured by the estimation of
zenith path delay and tilts of the refractivity field symmetry axis.
Formal uncertainties of group delays in quad-band experiments 
are at a level of 3--5~ps all year round, while the scatter of 
postfit residuals varies in a range of 15--25~ps. There should 
be some source of these errors that is greater in summer than in
winter. Path delay is greater in summer. If the atmospheric
path delay is entirely captured by the estimation process, 
an increase in wet path delay would not affect residuals. When 
estimating zenith path delay and tilts captures only \chalp{some}{a certain}
share of the total signal \addlp{that depends on schedule}, postfit
residuals \addlp{and the impact of the unmodeled atmospheric path delay}
will depend on the total path delay. The unmodeled \chalp{part}{share}
of path delay will affect estimates of other parameters, such as EOP.
The evidence we presented supports this explanation.

  Seasonality in errors is the most profound when most of the network
stations are located in mid-latitudes in the same hemisphere. 
Observations at a network that includes equatorial stations 
(no winter) or stations in different hemispheres (winter at one 
end of a baseline and summer at the other end) will exhibit less 
seasonality.

\subsection{EOP accuracy as a function of an observing session duration}

   Our extensive analysis of the impact of observing session duration on
EOP estimation revealed an interesting pattern from the comparison of vo 
and r1r4 experiments. When we increase session duration from 1 to 2--4 
hours, the EOP accuracy improves with the power law exponent 
\mbox{[-0.43, -0.46]}. After reaching the breaking point, the rms of 
differences decreases with the power law exponent \mbox{[-0.28, -0.30]}, 
much lower than the power law of -0.5 that describes the dependence of
 uncertainties of estimated parameters in the presence of uncorrelated 
noise. We treat this power law as a strong empirical evidence of the 
presence of correlation in noise. 

   What is the origin of this correlated noise? We consider the noise in 
group delays that contributes to right-hand sides of observation equations
has three components: 1)~thermal noise in the receiver; 
2)~instability of path delay in the VLBI signal chain hardware; 
3)~contribution from the unmodeled phenomena, either deterministic
or stochastic. The first contribution is due to radiation from individual 
molecules of receivers, which number is comparable to the Avogadro number 
($6 \cdot 10^{23}$) and therefore, is totally uncorrelated. In a similar
way, hardware errors are expected to be weakly correlated. But unmodeled
phenomena are expected to be correlated. The power law exponent
\mbox{[-0.28, -0.30]} implies that the overall contribution of all noise 
sources combined is highly correlated. As we saw, estimation of the 
atmospheric path delay in zenith direction retrieves the residual path 
delay only partly. The remaining path delay propagates to the right hand 
side. Considering that EOP errors have a strong seasonal variability, 
we connect these unmodeled phenomena with the contribution of residual 
atmospheric path delay that was only partly captured by the estimation 
process.

  When we injected an uncorrelated noise with a different variance, the 
power law is by modulo increasing, approaching -0.5 with an increase of 
the variance of the injected noise. This confirms that the power law of 
the dependence of EOP accuracy on session duration is indeed due to 
properties of the dominant noise contribution: when we added uncorrelated 
noise, we diluted the intrinsic noise. 

  The power law of the dependence of EOP accuracy on session duration is 
noticeably different for EOP derived from vo/r1r4 data and derived from
ca/na data. Two factors play a role. Firstly, the average reported 
uncertainty of group delay from vo experiments is 4~ps, while it is 20~ps 
for other observing sessions. That implies that the contribution of 
the thermal and hardware noise that is weakly correlated is 
minimal, and the share of the atmospheric noise is maximal. Our test
with injection of an additional uncorrelated noise suggests that this 
factor contributes. Secondly, the average number of observations in 
vo 24-hr experiments is 9110, compared with 4328 observations in 
r1r4 experiments, 1204 observations in ca campaign, and 988 
observations in na campaign. Our results indicate that the power 
law exponent depends not only on session duration, but on the total 
number of observations: the more observations, the flatter the power law.

  The fact that the dependence of EOP accuracy on segment duration
cannot be described by a single power law implies that there is another
factor that impacts accuracy of EOP derived from processing segments 
that are shorter than 2--4 hours. We should note that these experiments 
were designed as 24~hour blocks, and we artificially split
them into shorter blocks. When a segment becomes too short, separation
of estimated parameters, such as EOP, clock functions and atmospheric 
path delays in zenith direction is getting worse. Observations at high
and low elevations need to be present in a segment for a reliable 
estimation of atmospheric path delay. A schedule optimized for a 24-hr
block guarantees that such observations are present in a 24-hr block,
but it does not guarantee that such observations are present in any
1, 1.5, 2, or 2.5~hour long blocks. As a result, atmospheric path delay 
errors affect EOP estimates from processing shorter segments to 
a greater extent than they affect EOP from processing longer segments.

  Our results of processing 0.5~hr segments add an additional argument.
The rms of $E_3$ angles decreased by a factor of 1.09 ($2^{-0.13}$)
when the segment duration increased by a factor of 2. At the same 
time, the rms of $E_3$ angles decreased by a factor of 1.37 
($2^{-0.45}$) when segment duration increased by a factor of 2 from
1~hour to 2~hours in vo/r1r4 observing campaigns.  We attribute this
discrepancy to differences in scheduling strategy: vo/r1r4 were not
designed to run as short segments, but 1-hr experiments were.

  VLBI data that were used for deriving angles $E_3(t_1)$ and $E_3(t_2)$ 
from 0.5~hr blocks at $t_1$ and $t_2$ epochs are independent. The 
thermal noise in these data is independent and uncorrelated. If the 
only contribution to the noise were the thermal noise in receivers, 
then rms of the difference in $E_3$ estimates, $\sigma_0$, would have 
been $\sqrt{2} \, \sigma_0$  of the rms of individual $E_3$ 
estimates from 0.5~hr segments, i.e. 
$\sqrt{2} \: \cdot \: 1.42 = 2.01$~nrad. But we got $\sigma=1.47$~nrad, 
which is substantially less. We can compute correlation $\rho$ between 
estimates of $E_3$ angles derived from 0.5~hr date segments using 
a simple expression
\beq
    \rho = 1 - \Frac{1}{2}\Frac{\sigma^2}{\sigma_0^2}.
\eeq{e:e6}

  This gives us $\rho=0.46$. We consider this correlation as a direct
evidence of presence of a non-thermal noise due to mismodeling 
atmospheric path delay. Atmospheric path delay between two 0.5~hr time 
intervals is not independent and is correlated. Correlation in the 
atmospheric path delay between two segments causes correlation in 
estimates of $E_3$ angles determined from data of these segments. 
We should note that the correlation in the atmospheric path delay does 
not have to be the same as the correlation in $E_3$ estimates.

\iftrack \newpage \fi 
\subsection{Impact of source structure}\label{s:stru}

\addlp{
  \citet{r:and18} processed a 15~day VLBI observing campaign CONT14,
analyzed group delay misclosures over each triplet of stations, and
found that among 72~observed sources, 6 had small rms of group delay
misclosures, less than 10~ps. The rms of postfit residuals
over this group of sources was 19.2~ps versus 25.2~ps over all the
sources. Based on this finding, the authors  made a conclusion that
source-structure-related errors in geodetic VLBI are about as large
as all other error sources combined. However, if two processes have 
a similar impact on residuals, this does not necessarily mean that 
their impact on EOP estimates is the same. Here we make a rough 
estimate of the source stucture contriubution on EOP determination.

  Since we do not have an adequate model for source structure,
we can evaluate the impact of source structure contribution only
indirectly. Simulations of \citet{r:pla16} and our prior work for
determining source position using available source images \citep{r:gaia3}
demonstrated that source structure affects source positions mainly
along jet directions. 

  The contribution of structure contribution causes a jitter in source
position estimates. That process can be characterized by the first two
moments $\mu$ and $\sigma^2$. We should note that the first moment $\mu$
is absorbed in source position estimates and therefore, does not
impact EOP estimates.

  In \citet{r:radf} we estimated the rms of source position differences
along the jet direction derived from observations at 24~GHz with
respect to positions derived from dual-band 8.4/2.3~GHz observations:
0.60~nrad. We consider this as an estimate of the first moment of the 
source position jitter $\mu$ for the entire population of observed 
sources. We attribute these differences to the contribution of source 
structure. We consider with some reservations this estimate as a 
measure of the impact of source structure on source positions
on average. Although the differences in source positions from two bands
are not entirely from source source structure, which would cause an
overestimation of the impact, and source structure at two bands may be
in part common, which would cause an underestimation of the impact, this
is the best estimate of the impact based on evidence that we have now.

  Only a portion of source structure contribution propagates to
source position estimate, the remaining contribution propagates to
the residuals. But we assume here that only that portion that affects
source position may affect EOP estimates.

  If source position offset along the jet direction were the Gaussian
process, the first two moments $\mu$ and $\sigma^2$ had been independent.
But that would mean that for a given source position change due to
source stucture would occur in both directions along the jet with
an equal probabiulity, i.e. jets would randomly change sign of their direction.
Observations showed \citep[see, for instance,][]{r:mojave_xvii} that this
never happens. For one-side distributions, moments are not independent.
Since we do not know what is the actual distribution of source position
variations due to source structure change, following the Occam principle,
we consider two simplest distributions, the half-normal and Rayleigh, which
correspond to distributions of $|X|$ and $\sqrt{X^2 + Y^2}$, where $X$, $Y$
are normally distributed zero mean variables. The ratio of moments $\sigma/\mu$
for these distributions is $\sqrt{ \frac{\pi-2}{2} }$ and
$\sqrt{ \frac{4-\pi}{\pi} }$, respectively, or 0.60 and 0.76. This estimates
allow us to derive the upper limit of the variance in source positions due
to the unaccounted source structure contribution at 8.4~GHz: 0.4~nrad.
}

\addlp{
   To quantify the magnitude of the source structure contribution on EOP,
we performed a simulation. We added the random noise to source positions
with the zero mean and rms $1$~nrad along random uniformly distributed
directions. The seed of the random noise was reset for each observing
session. We found that a jitter in source position with a given rms $j$
affects the vector EOP estimates from r1r4 and vo networks as
$(0.32, 0.23, 0.37) \cdot j$. These coefficients depend on the network and
VLBI schedule. This allows us to provide a rough estimate of the impact
of source structure on EOP: $\sim\! 0.1$~nrad.

  Considering that the EOP accuracy from 24-hr experiments is 0.4--0.6~nrad,
we conclude that the impact of source unmodeled structure is not measurable.
}

\subsection{Formal and informal errors: what did we learn?}\label{s:form_errs}

  Our study confirmed what was already known for two centuries:
uncertainties from ground astronomical observations computed using 
the error propagation law are wrong. But why? Gauss-Markov theorem 
states the condition when uncertainties derived by the formula of 
the error propagation law (that is why they are called formal) are 
the dispersion of estimated parameters. We have no doubts in the 
proof of Gauss-Markov theorem. Therefore, we interpret a digression 
of our accuracy estimates from formal errors as evidence of 
a violation of conditions of the Gauss-Markov theorem. Namely, we 
conclude that the weight matrix we used in data analysis differs 
from the invert of the covariance matrix of noise.

  It was long known that atmospheric turbulence cannot be described 
as an uncorrelated random process \citep[see, for in 
instance,][]{r:tat71}. However, a quantitative assessment of the impact 
of an incorrect stochastic model used in data analysis requires 
significant resources for acquiring redundant data and their analysis. 

  We see that formal errors derived from data analysis using weight 
matrices with zero off-diagonal terms, which is equivalent to an assumption 
of no correlation, cannot be used as a measure of errors in EOP estimations
and cannot be used as a basis for optimization. Judging on formal errors,
one can conclude that construction of new antennas at Wettzell and Kokee 
Park, installing next generation quad-band recording system improved the
accuracy of $E_3$ angle by a factor of 2.1. Based on this metric one
could make certain recommendations for the future development aimed at 
improving the EOP accuracy. However, considering the achieved change in 
accuracy, 15\%, one could make rather different recommendations.

  Our analysis demonstrated that at networks of size of 0.6--0.8 Earth 
diameter, atmospheric conditions are the primary factors that affect accuracy, 
other factors as bandwidth, number of observations per hour being secondary. 
We interpret it as a manifestation of the dominance of atmospheric noise.
Our tests with an injection of the random noise demonstrate quantitatively 
that it is sufficient to reach a formal uncertainty of 20~ps of group delay 
determination for geodetic applications of VLBI techniques. Further
improvement in group delay uncertainty does not buy us anything. For 
instance, EOP accuracy from quad-band vo observations with 4~ps group 
delay precision turned out even {\it worse} than EOP accuracy from 
dual-band r1r4 observations with 20~ps group delay precision. 
That does not necessarily mean that a change in VLBI hardware 
caused degradation in EOP accuracy. We should note the vo network was 
evolving in 2019--2024, and the scarcity of meridional baselines in the 
first part of the interval was the factor that affected EOP estimates.

  \addlp{In order to see how the use of additive weights affected EOP accuracy,
we ran a solution without weight updates. The accuracies increased by
factors of 1.42, 1.23, and 1.41 for $E_1$, $E_2$, and $E_3$ components
respectively.}
  As we see, the use of the additive weight update did not solve the 
problem of underestimation of $E_3$ angles following the error 
propagation law, but only alleviated the problem. \chalp{\citet{r:lua08} 
proved a theorem stating that least squares with a full a~priori 
covariance matrix is equivalent to least squares with a diagonal weight 
matrix if and only if certain conditions are met, and weights and 
observation equations that we use in our solutions do not satisfy 
these conditions.}{\citet{r:lua08} have proven a theorem that
established the necessary and and sufficient conditions of equivalence
of a least square solution with a diagonal weight matrix with
a solution with a full a~priori covariance matrix: if there are 
$p$ linear combinations of columns of observation matrix that are 
eigenvector of $(\Cov(\epsilon,\tra{\epsilon}) W)$, where $p$ is 
the number of parameters, $(\Cov(\epsilon,\tra{\epsilon})$ is
the covariance matrix of right-hand sides and $W$ is the diagonal
weight matrix. It follows from the theorem that an update of diagonal 
elements is not sufficient to satisfy condition.
}
  
\subsection{Observations of different durations}\label{s:dur}
 
  If the dependence of EOP accuracy on session duration were 
described by the power law -0.5, the distribution of long and short 
VLBI observing blocks over a given interval would not have mattered
The broken power law of the dependence of EOP accuracy implies that 
planning observing sessions longer than 2--4~hours we get 
a diminishing return on used resources. Extending observing sessions 
from 4 to 24 hours, we have to spend 6 times more resources, but we 
gain only 70\% in accuracy instead of 140\%, because the atmospheric 
turbulence added a correlated noise. \remlp{Adding more stations does not 
improve noticeably the accuracy of $E_3$ determination.}

  If $E_3$ were the only valuable data product of geodetic VLBI, then we 
recommend to limit observations to one long baseline of 0.6--0.8 Earth 
diameter but restrict observing session durations to several hours, and 
to run them several times a day. Considering other data products, 
the recommendation would change, however, collected evidence shows that 
running experiments for 24 hours for EOP time series estimation is far 
from optimal.

\subsection{Similar comparisons made by others}\label{s:oth}

  An idea of evaluation of $E_3$ accuracy from 24-hr and 1-hr VLBI 
experiments is not new. A commonly used approach is to compare $E_3$ 
against IERS~C04 or similar combined series. For instance, \citet{r:haa21} 
compared $E_3$ derived from twelve 1-hr experiments spanned over three 
months against the USNO Bulletin~B. They presented rms of the differences 
from 11 solutions in a 1.7--2.3~nrad range (their Table~2). 
\citet{r:scha22} made a comparison of $E_3$ angle estimates from several 
1~hr Intensive experiments are reported rms differences against JPL~EOP2 
and IERS~C04 time series (their Tables 1 and 2) in a range of 
1.3--5.6~nrad. \citet{r:wan12} computed the rms of differences in $E_3$ 
angles determined from 1-hr observing sessions at KkWz baseline and 
IERS~C04 experiments over 2000--2011: 1.86~nrad. \citet{r:yao23} compared 
$E_3$ derived from 61 1-hr observing sessions at baselines KkWz and K2Ws 
in 2021. They made comparison of $E_3$ angles with respect to IERS~C04 and 
reported rms of the differences: 2.05 and 1.83~nrad. In addition, they 
split 1-hr K2Ws into two 0.5~hr long blocks in the beginning and the end 
of an 1-hr observing session and processed them individually.
\citet{r:yao23} report that the rms of differences with respect to 
IERS~C04 increased by 7\% when 0.5-hr blocks at the beginning of the 
observing session were processed and decreased by 10\% when 
30~minute blocks at the end of the observing sessions were 
processed. According to the F-test, these differences are insignificant
at the 0.1 confidence level. That means that throwing away 1/2
of K2Ws data did not change results. This finding contradicts to the
broken power law that we derived by thinning 24-hr datasets.
Inspired by these results, we performed our analysis. Our analysis
showed that throwing away half of the data degraded accuracy, but 
only at 9\% instead of 41\% if the noise were uncorrelated.

  \citet{r:raut22} compared $E_3$ angle estimates determined from
processing 1-hr and 24-hr experiments at three independent campaigns 
during 15 days in November/December 2017. They performed a number 
of solutions, including splitting 24-hr experiments into 1-hr blocks
that they called pseudo-intensives. They report in Table~5 the rms 
of differences with respect to IERS~C04 1.97~nrad for 24-hr blocks
and 1.85~nrad for 1-hr blocks. This finding also does not support the
broken power law we found, but the rms of the differences are 
substantially greater than in our results.

  All these comparisons are made with respect to $E_3$ time series 
derived from other VLBI results with applying smoothing, which makes 
their interpretation difficult since three factors affect the 
differences, 1)~intrinsic errors of $E_3$ from 1-hr experiments; 
2)~intrinsic errors of errors of $E_3$ that were used for generation 
of derived time series; and 3)~errors introduced by smoothing.

  For completeness, we mention that \citet{r:gip16} 
published results of a campaign of 9 experiments with a design similar 
to our campaign s22. According to the campaign description, each 
experiment contained twenty two 1-hr scheduling blocks at KkWz baseline 
among with data at other stations. However, database files in the public 
archive at NASA Crustal Dynamics Data Information System (CDDIS) contains 
data at all stations, except stations {\sc kokee} and {\sc wettzell}.
Apparently, \citet{r:gip16} based their analysis on additional 
datasets that were not publicly released. Considering governing open 
science policies, we defer comments on these results till missing 
datasets will become available. 

  \citet{r:mcm22} presented wrms of the differences of EOP from
two concurrent VLBI networks during 15 consecutive days during
CONT17 observing campaign in November/December 2017. He reports the 
rms of the differences in Euler angle vector (0.21, 0.19, 0.25)~nrad. 
This is a factor of 2--3 smaller than the rms of the differences
in EOP derived from analysis of group delays from vo/r1r4 experiments
presented in Table~\ref{t:conc_24hr}. There are three factors that 
played the role. First, CONT17 experiments are not representative.
This campaign was an attempt to get the best results using existing
VLBI hardware, while vo and r1r4 experiments have been running as
regular, operational observing sessions. Second, these experiments
ran in winter time for most of the stations in the network. And third, 
statistics over a continuous 15~day period are different from
statistics over a 5~year period in a presence of red noise. 
\citet{r:ray17} argues strongly that the longer the time period
for error characterization, the greater the errors, and we find
their argumentation quite convincing. \citet{r:rfc1} presented strong 
evidence of the impact of red noise on results of comparisons of 
two subsets of VLBI data on source positions. At present, we are not 
in a position to deliver a quantitative proof of the presence of red 
noise in EOP estimates at scales of 0.05--5~years, but we consider 
that the first two factors are not sufficient to explain the 
discrepancies in statistics of that magnitude.

\subsection{Comparison with IERS C04 time series}\label{s:iers}

   \addlp{As we see from the previous section,} it is customary to 
compare EOP estimates against IERS~C04 series \citep{r:iers_c04},
\addlp{implicitely assuming that C04 represents the groud truth,
and therefore, the differences against C04 can be considered as
a measure of accuracy.} \chalp{Therefore,}{Since the reported 
statsitics of the differences with respect to IERS~C04 disagree
with our estimates of EOP accuracy based on processing concurrent
observations}, it is instructive to compare our 
\chalp{estimates of EOP accuracy from analysis of concurrent VLBI experiments}
{EOP estimates} against the C04 time series. The time series of $E_1$ 
and $E_2$ angles in C04 time series are dominated by the IGS final data 
product from processing independent GNSS phase and pseudorange data. Therefore, one 
can expect that the statistics of the rms of differences against C04 will 
be close to our estimates of accuracy based on the three corner method. 
Indeed, it turned out very close for vo campaigns, but 13--17\% higher 
for r1r4 campaigns. Time series $E_3$ in C04 are derived from the results 
of processing the same VLBI group delays. According to \citet{r:iers_c04},
the time series of $E_3$ angles are derived from a)~$E_3$ estimates from 
analysis of R1 and R4 VLBI experiments by 10~analysis centers and then 
averaged with some weights by the so-called International VLBI Service 
for Geodesy and Astrometry (IVS) combination center; b)~$E_3$ 
estimates from some 1-hr programs: $4\times$ daily from the Institute of 
Applied Astronomy in Sankt-Peterburg, Russia, and several programs
processed by the United States Naval Observatory (USNO) on days when 
there is no r1r4 experiment. The input time series are then smoothed by 
using the low-pass filter. Therefore, the IERS C04 accumulates errors 
due to deficiencies in data analysis of individual groups and 
suppresses, to some degree, fast variability in EOPs due to smoothing.

  Both the differences of EOP estimates against IERS C04 and statistics
of the difference from concurrent VLBI experiments provide measures
of EOP accuracy. It is instructive to compare them. 
Table~\ref{t:eop_acc_c04} shows these accuracy estimates. 

\begin{table}
   \caption{The estimates of EOP accuracy from the comparison of 
            concurrent VLBI observations (column~3) and 
            the rms of differences of EOP derived from our VLBI 
            solution against IERS C04 (column~4). The estimates
            in column 3 for the first four rows are derived from 
            the three-corner hat method, and for the last for
            rows are derived from comparison of EOP from concurrent
            observations.
           }
  \begin{tabular}{l r c c}
        \hline
        Comp          & \ntab{c}{duration} & \ntab{c}{concur} & \ntab{c}{IERS C04} \\
                      & \ntab{c}{hour}     & \ntab{c}{nrad}   & \ntab{c}{nrad}     \\
        \hline
        $E_1$ vo        & 24 & 0.46    &  0.52  \\
        $E_2$ vo        & 24 & 0.57    &  \chalp{0.67}{0.72}  \\
        $E_1$ r1r4      & 24 & 0.34    &  \chalp{0.34}{0.39}  \\
        $E_2$ r1r4      & 24 & 0.32    &  \chalp{0.34}{0.42}  \\
        $E_3$ vo/r1r4   & 24 & 0.40    &  \chalp{0.77}{0.75}  \\
        $E_3$ K2Ws      &  1 & \chalp{1.30}{1.40} &  1.52  \\
        $E_3$ MkWz      &  1 & \chalp{1.21}{1.32} &  1.60  \\
        $E_3$ KkWz      &  1 & \chalp{1.50}{1.59} &  1.48  \\
        \hline
   \end{tabular}
   \label{t:eop_acc_c04}
\end{table}

  It turned out that the differences of $E_3$ angles from 24-hr experiments
are a factor of \chalp{1.7}{1.8} greater than the accuracy of these angle
estimates derived from our processing concurrent vo/r1r4 experiments
(compare columns 3 and 4 in Table~\ref{t:eop_acc_c04}). In general, 
the accuracy of a combined series like IERS C04 is governed by the accuracy 
of input series and smoothing. \chalp{We can rule out smoothing errors}{Smoothing
is not expected to be a major factor,} since 
the epochs of our EOP estimates and the EOP estimates of IVS combined series 
are practically the same. Therefore, we conclude that the IVS combined
$E_3$ angle series are a factor of 1.8 less precise than our results.
The differences of $E_3$ angles derived from 1-hr experiments with respect
to the C04 series are a factor of \chalp{1.17--1.32}{1.1--1.2} greater than
the accuracy of these angle estimates from concurrent experiments,
except for the time series from KkWz baselines.

  \addlp{We have a disagreement with comparisons made by others. In order to
understand the origin of this disagreement, we took the IVS combined EOP 
series\footnote{\web{https://www.ccivs.bkg.bund.de/data/QUAT/COMBI/ivs24q3e.eops}
accessed on May 16, 2025.} and} computed the statistics of the differences 
of IVS Combined EOP from 148 r1r4 experiments with respect to our solution 
after removal of the linear trend: the \chalp{rms}{standard deviations} of 
the differences were 0.31, 0.38, and 0.86~nrad for $E_1$, $E_2$, and $E_3$ 
respectively. Similar statistics for EOP derived from 148 vo experiments are 
0.61, 0.66, and 0.99~nrad. \addlp{Statistics of the standard deviations
in differences between IVS and IERS time series over 265 vo/r1r4 experiments 
are 0.56, 0.62, and 1.09~nrad.}

  To dig even deeper, we compared EOP estimates from the same list of 
concurrent vo and r1r4 observing programs using the IVS combined time 
series. The epochs of EOP in the IVS combined time series are close, within
2~hours, but not the same. We propagated the EOP on the common epoch
using provided estimates of EOP rates. The rms of the differences divided
by $\sqrt{2}$ are shown in Table~\ref{t:conc_comp_ivs}. We see that our 
solution provides the rms of the differences a factor of 1.4 smaller
for the $E_3$ angle. \chalp{That prompts us to interpret}{Therefore, we 
conclude that} the relatively large differences in IVS Combined EOP 
\addlp{time series} with respect to our solution as a deficiency of the 
IVS Combined series. 

\begin{table}
   \caption{The estimates of EOP accuracy from comparison of concurrent
            VLBI vo and r1r4 observing campaigns: our solution 
            (the 2nd columns) and the IVS combined EOP time series 
            (the 3rd column). The rms of the differences is divided 
            by $\sqrt{2}$.
           }
   \begin{tabular}{l  r r  @{\qquad\quad} r r}
        \hline
        Comp          & \nntab{c}{Our solution} & \nntab{c}{IVS Combined} \\
                      & bias & rms  & bias & rms  \\
                      & nrad & nrad & nrad & nrad \\
        \hline
        $E_1$ &  \addlp{0.20} & \chalp{0.52}{0.54} &  \addlp{0.60}  & 0.68  \\
        $E_2$ &  \addlp{0.51} &        0.71        &  \addlp{-0.44} & 0.69  \\
        $E_3$ &  \addlp{0.14} & \chalp{0.44}{0.40} &  \addlp{0.51}  & 0.57  \\
        \hline
   \end{tabular}
   \label{t:conc_comp_ivs}
\end{table}

\iftrack \newpage \fi 
\remlp{\subsection{IVS Combined EOP time series and IERS C04}} \label{s:comb}

\remlp{
  We established earlier that the IVS Combined EOP series has a notable
deficiency. Since the time series from r1r4 experiments are used for 
IERS~C04, it inherits this deficiency. We should note that the IVS 
combined solutions fall into the category of operational solutions, 
while our solutions fall into the category of re-processing. We base
our analysis on 50~years of efforts to improve the VLBI data analysis 
technique pursued by NASA Goddard Space flight center, and our data 
analysis technique differs from the technique commonly used by other 
groups for operational data analysis. In particular, we see no 
statistically significant biases,
no systematic patterns when we compare EOP from 24-hr and 1-hr 
experiments or EOP from different observing sessions.

  Computing a combined solution, i.e. the weighted mean of results, may
have some advantage for operational time series because it helps to detect
a failure in quality control of an individual analysis center and 
allows us to mitigate situations when a given analysis center is 
offline. On the other hand, processing the same dataset several times and 
then averaging results contradicts Gauss-Markov theorem, and that causes
a deviation of such a result from the optimum.

  We should bear in mind worse accuracy of a combined data series,
treat it as a temporary, coarse solution, and avoid using it for 
science. IERS~C04 time series falls into the same category of operational 
data products since it is based on operational solutions. We have 
demonstrated that EOP estimates from a uniform reprocessing using advanced 
theoretical model and parameter estimation technique of all accumulated 
data are superior. In a similar way, campaigns of reprocessing GNSS 
observations with an improved model \citep[see, for instance,][]{r:dach21} 
also provide superior results.
}

\section{Summary and Outlook}\label{s:sum}
 
\begin{figure}
   \noindent
   \centerline{\includegraphics[width=0.61\textwidth]{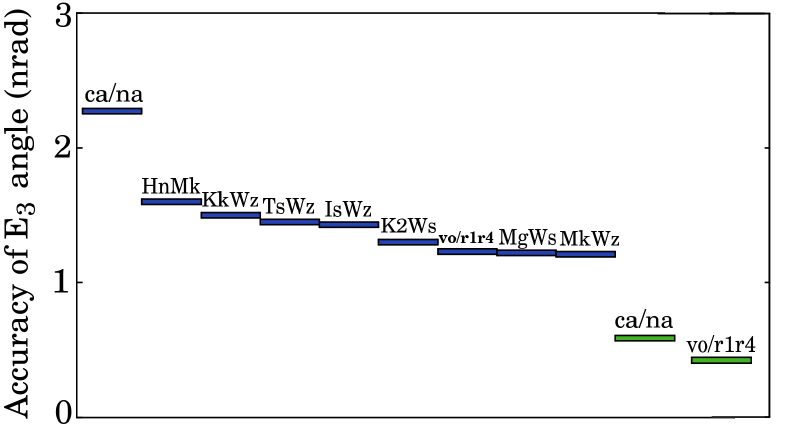}}
   \caption{Accuracy of $E_3$ angle estimates derived from analysis of 
            different concurrent experiments. The blue thick lines 
            (leftmost 9) show accuracy from 1-hr experiments or 
            processing block. The green thick lines (rightmost 2)
            show accuracy from 24-hr experiments.}
   \label{f:eop_acc}
\end{figure}

   Having processed a number of recent and historical datasets of 
concurrent VLBI observations at different networks, we have arrived 
at more reliable estimates of the accuracy of time variable Euler 
angles describing the Earth's rotation. These estimates have a little 
resemblance to formal computed using the law of error propagation 
under the assumption that the noise is uncorrelated. \addlp{They
also noticeably smaller than the differences against IERS~C04 time 
series that was often used as a measure of EOP accuracy in literature.}
We summarize the accuracy of determination of $E_3$ angles from various 
24-hr and 1-hr campaigns in  Figure~\ref{f:eop_acc}.

   We confirmed that the accuracy of EOP estimates from VLBI observations 
cannot be characterized by formal errors. Accuracy is determined 
mainly by the network and atmospheric processes. 

   We found strong evidence that the dominant error source that determines
accuracy of geodetic VLBI results is the residual atmospheric path delay, 
which is variable and correlated. The dominance of the atmospheric noise 
manifests itself in a form of a)~seasonality in EOP accuracy --- winter 
results are accurate at the high confidence level; b)~dependence of 
accuracy on experiment duration as the power law -0.3; c)~the insensitivity 
of results on the precision of group delay when it drops below the 20~ps 
level; and d)~correlation 0.46 between estimates in $E_3$ angles computed
from independent group delay datasets 0.5~hours apart. 

  \addlp{Based on statistics of differences in positions of a large sample 
of sources determined from analysios of observations at different bands 
reported in literature, we got the quantative estimate of the upper limit 
of the impact of source structure on EOP estimates: 0.1~nrad. Comparing the 
extra variance in EOP estimates in summer versus in winter with the impact 
of source structure on EOP, we conclude that the impact of source structure 
contribution is one order of magnitude less than the impact of unmodeled 
atmospheric path delay.}

   We found no evidence that advanced scheduling strategies for
improved determination of the atmospheric path delay in zenith direction
have any impact on the accuracy of EOP estimates. Inability to predict the 
accuracy based on simulation suggests that our approaches to data analysis 
and simulation of geodetic VLBI have to be revised. We recognize that this 
inability poses a serious problem. \addlp{\citet{r:tiege25} simulated 
an increase of repeatabilities in EOP estimates when 1/4 observations is 
removed, and they found that $E_3$ errors defined as repeatabilities increased 
by 18.8\%. In contrast, our analysis of real data showed that $E_3$ accuracy
from 18~h experiments differs by less than 1\% with respect to the accuracy
derived from analysis of 24~hr experiments.}

  The use of a weight matrix with filled diagonal terms and zero off-diagonal 
terms was a necessity in the 19th and 20th centuries due to the scarcity 
of computing resources, but the use of this technique by inertia in the 21st 
century is an anachronism. We are going to enhance our data analysis procedure 
by the use of a full weight matrix, including off-diagonal terms, and 
encourage others to pursue these efforts. The goal of these efforts is 
to predict accuracy of EOP estimates within tens of percent, explain the power 
law -0.3 and seasonality in EOP accuracy, and explain correlation 0.46 between 
estimates of $E_3$ angles from 0.5-hr data segments. The dataset presented
in this study is an excellent testbed that can be used for tuning 
algorithms that eventually are supposed to produce correct estimates of 
uncertainties of estimated parameters in the presence of a correlated noise.

  The dependence of EOP errors as a power law of -0.3 for VLBI experiment
durations longer than 4 hours and insensitivity of estimates of $E_3$
angle to the number of observing stations raises the issue of optimization
of the program of geodetic VLBI observations. The current strategy of
allocation of very significant resources for observing 24-hr sessions 
at a 10--12~station network with long gaps between sessions does not favor 
$E_3$ angle determination. The optimization of the observing strategy 
based on the power law of the dependence of EOP accuracy on session 
duration is a topic for a future research.

  Although we have arrived at our conclusions based on the analysis of 
VLBI observations, the same atmospheric noise affects GNSS observations
and therefore, our findings can be generalized to other microwave 
observations.

  \addlp{We strictly adhered in our research SI units adopted in 1960.
In order to facilitate perceiving our results by those readers who 
got used to units that are not a part of the international system, 
we translate some important results to commonly used units. We found that 
accuracy of polar motion is 68~$\mu{as}$ from r1r4 experiments and 
106~$\mu{as}$. Accuracy of UT1 angle from r1r4/vo observations is 
5.5~$\mu$sec. Accuracy of UT1 from long time series of 1-hr observations 
is 19--22~$\mu$sec. Accuracy of UT1 from dedicated s22 observations varies 
from 9~$\mu$sec (winter) to 20~$\mu$sec (summer).}

\bmhead{Acknowledgements}
 
  We would like to thank Frank Lemoine, \addlp{Christian Bizouard, and Chris 
Jacobs} for valuable comments. It is our pleasure to thank staff at the CDDIS 
for making available Level 2 VLBI data. We thank Cynthia Thomas and Dirk 
Behrend for coordinating International VLBI for Geodesy and Astrometry 
observing programs. \addlp{We also would like to express our gratitude
to the staff of observing geodetic stations, VLBI correlators, technology
development groups, and data centers working together and forming IVS for 
their dedication. The overview of IVS can be found in \citep{r:ivs}}.

\section*{Declarations}
 
 \begin{itemize}
     \item Funding: L.P. was supported by the NASA Space Geodesy Project.

     \item Conflict of interest/Competing interests: none

     \item Code availability: we used for our analysis open source software 
           package SGDASS available at 
           \web{https://doi.org/10.25966/7z5f-eg07}. 
           The s22 schedules are generated using the VieSched++ software 
           package \citep{r:scha19} available at 
           \web{https://github.com/TUW-VieVS/VieSchedpp} under the GNU 
           General Public License.

     \item Materials availability: {\it all} Level 2 VLBI data used in our 
           analysis (group delays) are publicly available at 
           \web{https://cddis.nasa.gov/archive/vlbi/ivsdata}

     \item Author contribution: L.P., M.S., and C.P. designed s22 observing 
           campaign; M.S. developed s22 experiment schedules; C.P. ran 
           experiments, processed Level~0 and Level~1 data; L.P. ran data 
           analysis of Level~2 data and performed statistical analysis of 
           results. All authors discussed the results, commented on the 
           manuscript, and agreed to be held accountable for the work.
     \item Use of artificial intelligence: we certify that no machine 
           learning or artificial intelligence techniques were 
           used, neither during data analysis, nor in manuscript preparation. 
\end{itemize}

\iftrack \newpage \fi 

\begin{appendices}
\section{Scheduling strategies}\label{a:sched_22}

\begin{figure}
    \centering
    \includegraphics[width=0.75\linewidth]{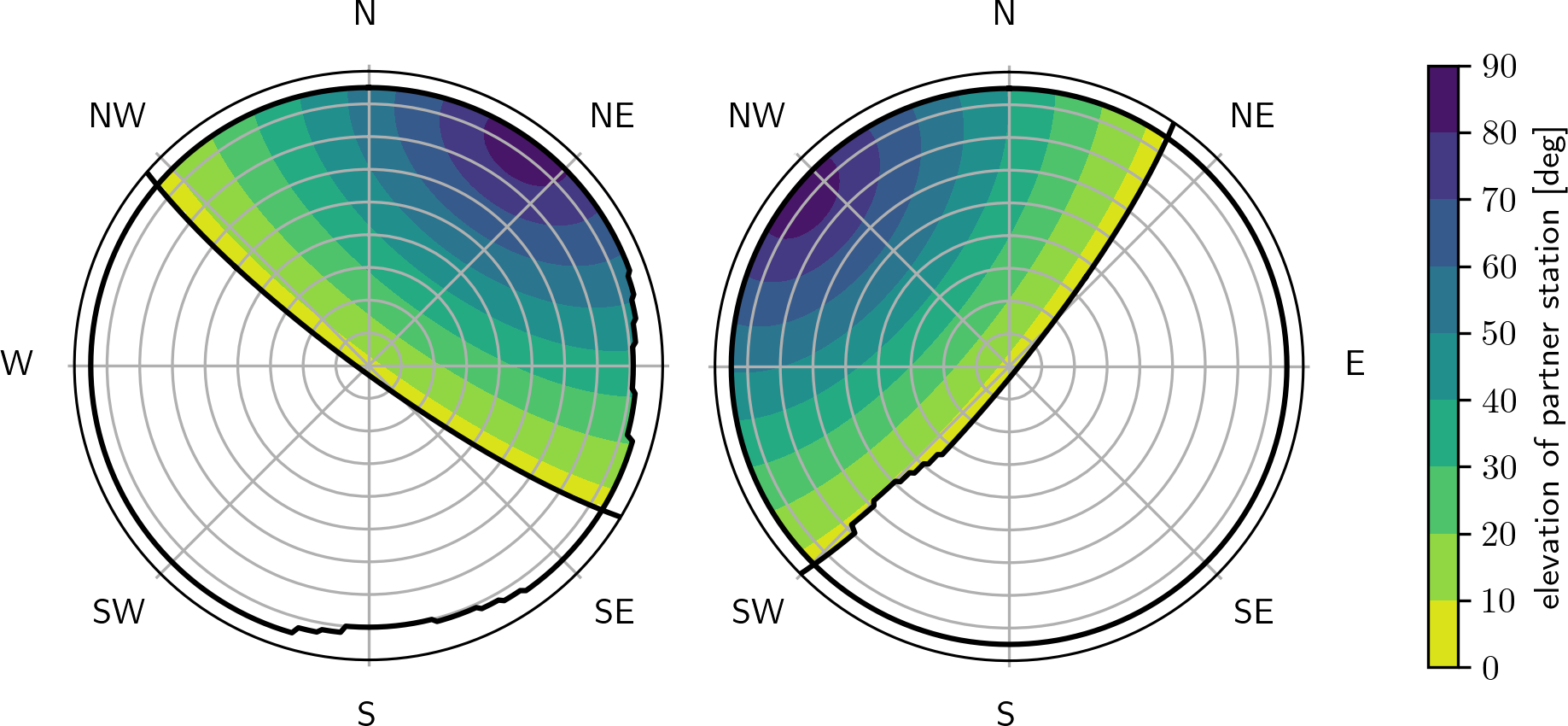}
    \caption{Mutual visibility between \chams{Ws}{Mg} (left) and \chams{Mg}{Ws} (right) stations. 
             The elevation of the other station is color-coded. 
             The black line marks the visible horizon \citep{r:scha23}. 
             }
    \label{f:scheduling_el}
\end{figure}

\addms{
  While generally speaking, observations at longer baselines are more 
sensitive for measuring Earth's rotation angle $E_3$, they also face 
limitations due to a reduced common visibility of the sky
\citep{r:scha21}. The $\sim\! 8400$~km baseline between stations Mg and 
Ws presents certain geometric advantages. 
At this distance, observations near the zenith at one station correspond 
to observations at very low elevation angles at the other station 
(see Figure~\ref{f:scheduling_el}). This geometry is particularly 
beneficial for solving for residual atmospheric path delay in zenith
direction, which remains the primary source of error in VLBI measurements. 
To effectively separate atmospheric zenith path delay from other 
parameters, such as station height and clock biases, it is essential 
to observe at a wide range of elevation angles.
}

\addms{
  The s22 sessions were specifically designed to investigate improved
estimation of residual atmospheric path delay in zenith direction.
Each session is divided into 1-hour blocks that alternate between two 
scheduling strategies. Strategy~A follows the conventional scheduling 
approach utilized in the VieSched++ scheduling software 
\citep{r:scha19} and serves as a baseline reference. By contrast, 
strategy~B is tailored to maximize the variability in elevation angles, 
promoting frequent alternation between high- and low-elevation 
observations to improve sensitivity to zenith atmospheric path delay 
parameters.
}

\addms{
  To implement Strategy~B, sources at high elevation for each station are 
identified in advance. Scans are then scheduled according to the following 
repeating pattern: 

\begin{enumerate}
    \item a source at high elevation for Mg (corresponds to low elevation for Ws),
    \item any source,
    \item a source at high elevation for Ws (corresponds to low elevation for Mg),
    \item any source.
\end{enumerate}
}

\addms{
  The interleaved ``any source'' steps (2 and 4) help to maintain a balanced 
distribution of observations across different sources. In general, a minimum 
interval of 20 minutes is enforced between consecutive observations of the 
same source. However, due to the limited availability of high-elevation sources 
at each station, this restriction is relaxed to 10 minutes for those specific 
sources in strategy B. Additionally, to ensure adequate coverage of 
low-declination regions, a low-declination source is explicitly scheduled 
every 15 minutes in both strategies.
}
\addms{The impact of this strategy on the distribution of observation is discussed in \citep{r:scha23}.}

\section{Relationship between daily estimates of EOP in different notations}
\label{a:hist_eop}

\addlp{
  We used in our analysis a simple mathematical model in expression~\ref{e:e43} 
to describe dependence of time of residual Euler angles with Earth
Orientation Parameters $P_c, P_s, S_c, S_s, R_c, R_s, E_i(s), \dot{E}_i(s)$. 
An alternative mathematical model was introduced by \citep{r:her86a}:
}

\addlp{
\beq
   \vec{q}_e(t) = \left(
   \begin{array}{llll}
      \mu ( \, Y_p(s)  & + \dot{Y_p}(s) \cdot (t-t_{\rm ref}) & \hspace{-6.5em} + 
            \Delta X(s) \cos -\Omega_n t & + \Delta Y(s) \sin -\Omega_n t ) \vhx \\
      \mu ( \, X_p(t)  & + \dot{X_p}(s) \cdot (t-t_{\rm ref}) & \hspace{-6.5em} + 
            \Delta X(s) \sin -\Omega_n t & - \Delta Y(s) \cos -\Omega_n t ) \vhx \\
      \kappa  (\, {\rm UT1-UTC}(s)) & + \Frac{d}{dt}({\rm UT1 - UTC})(s) \cdot (t-t_{\rm ref})) &&   \\
  \end{array}
  \right),
\eeq{e:a1}
  where $X_p$, $Y_p$, polar motion, $\Delta X$, $\Delta Y$ are pole offsets,
and $\mu$ and $\kappa$ are scaling factors introduced in section \ref{s:for}.
}

\addlp{
  Comparing with expression~\ref{e:e43}, we immediately see that 

\beq
   \begin{array}{lcl}
      \mu Y_p(s) & \approx & E_1(s) \vx \\
      \mu X_p(s) & \approx & E_2(s) \vx \\
      \kappa ({\rm UT1-UTC})(s) & \approx & E_3(s) \vx \\
      \mu \dot{Y}_p(s) & \approx & \dot{E}_1(s) \vx \\
      \mu \dot{X}_p(s) & \approx & \dot{E}_2(s) \vx \\
      \kappa \Frac{d}{dt}({\rm UT1-UTC})(s) & \approx & \dot{E}_3(s) \vx \\
   \end{array}
\eeq{e:a2}
}

\addlp{
We put sign $\approx$ because the two parameterizations of harmonic 
variations are not equivalent. The model in expression \ref{e:a1}
is less precise. The differences in the first order approximation are 

\beq
  \displaystyle \Frac{1}{t_b - t_e} \, \int\limits_{t_b}^{t_e} \:
        \sum_{j}^{N} P^s_j \sin ((\omega_j + \Omega_n)\, t + \phi_j) \, dt,
\eeq{e:a3}
   where $t_b$ and $t_e$ are start and stop of an experiment. These
differences are small for spectrum constituents with frequencies
close to $-\Omega_n$, i.e nutations with long periods viewed in the
celestial coordinates system. Celestial pole offsets can be related
to $P^c,P^s,S^c,S^s$ as
}

\addlp{
\beq
   \begin{array}{lclcl}
      \mu \Delta X(s) & \approx & \displaystyle \sum P^c_j & + & (t-t_0) \cdot S^c \vx\vx \\
      \mu \Delta Y(s) & \approx & \displaystyle \sum P^s_j & + & \phantom{(t-t_0) \cdot\:\,} S^s, \\
   \end{array}
\eeq{e:a4}
   where summation is performed only over constituents of the spectrum
in the retrograde diurnal band. We should note that no $R^s, R^c$ terms are
accounted when parameterization in equation \ref{e:a1} used. If the
a~priori model of the dependence of Euler angles on time is known precisely,
the difference between this parameterization and the parameterization in
equation~\ref{e:e43} is negligible. However, errors of the a~priori model
will affect $E_1(s), E_2(s), E_3(s)$ and $X_p(s), Y_p(s)$, and UT1-UTC(s)
differently.
}

\end{appendices}

\bibliography{conc}

\end{document}